\documentclass[12pt,preprint]{aastex}
\usepackage{ulem}
\usepackage{color}

\usepackage{mathtools}

\newcommand{\rhocutoff}{\rho_\mathrm{cutoff}}
\newcommand{\rhoanelastic}{\rho_\mathrm{anelastic}}

\newcommand{\gcc}{\mathrm{g~cm^{-3} }}
\newcommand{\Tcutoff}{T_\mathrm{cutoff}}

\newcommand{\msolar}{\mathrm{M}_\odot}

\begin{document}

\title{Low Mach Number Modeling of Convection in Helium Shells on 
       Sub-Chandrasekhar White Dwarfs. I. Methodology}

\shorttitle{Sub-Chandra Convection. I. Methodology}
\shortauthors{Zingale et al.}

\author{M.~Zingale\altaffilmark{1},
        A.~Nonaka\altaffilmark{2},
        A.~S.~Almgren\altaffilmark{2},
        J.~B.~Bell\altaffilmark{2},
        C.~M.~Malone\altaffilmark{3},
        R.~J.~Orvedahl\altaffilmark{1}}

\altaffiltext{1}{Department of Physics \& Astronomy,
                 Stony Brook University,
		 Stony Brook, NY 11794-3800, USA}

\altaffiltext{2}{Center for Computational Sciences and Engineering,
                 Lawrence Berkeley National Laboratory,
                 Berkeley, CA 94720, USA}

\altaffiltext{3}{Department of Astronomy \& Astrophysics,
                 The University of California, Santa Cruz,
                 Santa Cruz, CA 95064, USA}

\begin{abstract}
We assess the robustness of a low Mach number hydrodynamics algorithm for
modeling helium shell convection on the surface of a white dwarf in the 
context of the sub-Chandrasekhar model for Type Ia supernovae.  
We use the low Mach number stellar hydrodynamics code,
MAESTRO, to perform three-dimensional, spatially-adaptive simulations of convection 
leading up to the point of the ignition of a burning front.  We show that 
the low Mach number hydrodynamics model provides a robust description of 
the system.
\end{abstract}

\section{Introduction}

Sub-Chandrasekhar models for Type Ia form an attractive progenitor
candidate because of the abundance of low mass white dwarfs.  The
modern model of these explosions traces back to the work of
\citet{livne:1990, livneglasner,
  woosleyweaver:1994,wiggins:1997,wiggins:1998} amongst others.  In
this ``double detonation'' model, a detonation ignited in the accreted
helium layer on the surface of a low mass carbon/oxygen white dwarf
drives a shock inward, compressing the star and initiating a
detonation in the core.  At the time of these calculations, a common
concern was that the lack of resolution in simulations prevented a
realistic investigation of these events.  Recently, however, the
sub-Chandrasekhar mass progenitor model of Type Ia supernovae has seen
renewed interest~\citep{fink:2007,fink:2010,sim:2010,shen:2010}, as
observations show an increasing diversity of SNe Ia events.

As in the traditional Chandrasekhar-mass model of SNe Ia, the ignition
of a burning front is preceded by a long period of convection.  Here,
however, the convection takes place in the He layer, driven by
reactions at the base of the layer.  To date, numerical simulations of
this type of explosion have been initialized by seeding a detonation
in the He layer.  However, it remains an open question as to whether
the turbulent convective flow in the He layer can ignite a detonation
in the first place; it is also quite possible that there can be
multiple nearly-simultaneous ignitions (see for example,
\citealt{garcia:1999}).

A potential issue with this model is that a detonation in the He shell
would produce large amounts of $^{56}$Ni at the edge of the star that
is inconsistent with observations
\citep{hoeflichkhokhlov:1996,hoeflich:1996} (also see
\citealt{nugent:1997} for additional concerns about the spectra).  The mass
of the He layer is uncertain, and it has been suggested that one way
to address the overproduction of Ni is for the He shell to be very
thin, in which case it may be able to detonate without over-producing
surface $^{56}$Ni~\citep{fink:2010,kromer:2010}. Detailed one-dimensional stellar
evolution calculations suggest that the helium in these smallest mass
shells may never ignite as a detonation to begin
with~\citep{woosleykasen:2010}.  If it does ignite,
\citet{townsley:2012} showed that a robust detonation can propagate
through the layer.

The dynamics of the convection and development of a burning front
in the sub-Chandrasekhar model is inherently multidimensional.
Because the fluid and flame velocities in this initial phase are much 
lower than the sound speed, we can apply the same low Mach number methodology 
used to study ignition in the Chandrasekhar-mass
scenario \citep{paper4,wdconvect,wdTurb}.  Once we ensure that we have 
a robust simulation methodology, there is a large parameter space of 
initial models to explore.  The goal of this first paper is to demonstrate that
low Mach number hydrodynamics provides an efficient and accurate
simulation platform to explore the convective stage of these sub-Chandra
events.

\section{Numerical Methodology}\label{sec:Numerical Methodology}

To study the turbulent convection in the sub-Chandrasekhar model,
we use the MAESTRO code as documented in~\cite{multilevel}.  
For an overview of the low Mach number equations and
numerical methodology, we refer the reader to Section 2.1 of \cite{wdTurb}.
In summary, MAESTRO is a finite-volume, adaptive mesh 
stellar hydrodynamics code that models the flow using a low Mach number 
approximation---sound waves
are filtered out of the system, but compressibility effects due to
stratification and local heat release are retained.  For most
of the simulations presented here, the star is modeled on a Cartesian grid
with the center of the white dwarf at the coordinate origin, i.e., we model
one octant of the full star.  Since the convection
is confined to the outer shell, away from the center of the star, this
geometry captures the convective behavior well.  The base state
pressure and density are represented by one-dimensional radial profiles
which capture the hydrostatic state of the star.  As in the simulations in
\cite{paper4,wdconvect,wdTurb}, we derive the temperature from the 
equation of state given the pressure, density and mass fractions, 
rather than evolving the enthalpy equation.
Below we provide additional details specific to the simulations presented here;  
these introduce a variety of parameters that will be explored in the results section.

\subsection{Microphysics}

We use a general, publicly available equation of state consisting of
ions, radiation, and arbitrarily degenerate/relativistic electrons,
together with Coulomb
corrections~\citep{timmes_eos,timmes_swesty:2000}.  We use a simple
reaction network consisting of the triple-alpha and
$^{12}\mathrm{C}(\alpha,\gamma) ^{16}\mathrm{O}$ reactions.  The rates
are from \citet{caughlan-fowler:1988}, with screening as in
\citet{graboske:1973,weaver:1978,alastuey:1978,itoh:1979}.  The
$^{12}\mathrm{C}(\alpha,\gamma) ^{16}\mathrm{O}$ reaction rate has
been multiplied by 1.7 as suggested by \citet{weaver-woosley:1993} and
\citet{garnett:1997}.  This network is an extension of the network used by
\citet{malone:2011}.  When computing the effect of reactions over a
time interval, we evolve the temperature along with the mass
fractions, keeping the thermodynamic derivatives frozen during the
integration, as described in~\citet{ABNZ:III}.

\subsection{Initial Model}

For simplicity, we construct our own one-dimensional, semi-analytic initial 
model of a white dwarf with a helium layer.  This allows us to control specific
features as we learn about the algorithmic sensitivity to the choice 
of initial model parameters.
The initial model is constructed with an isothermal C white
dwarf with an isentropic He layer on the surface.  When initializing
the data on the three-dimensional grid, we interpolate from this initial
model and then apply a velocity perturbation, the latter described in 
Section \ref{sec:velocity init}.

To construct the initial model, we use the following iterative process:
\begin{itemize}
\item We start by providing an estimate for the white dwarf central density,
  $\rho_\mathrm{core}$, and the density at which we transition to helium,
  $\rho_\mathrm{He}$.  We also specify the composition of the core,
  $X_\mathrm{core}$, and its temperature, $T_\mathrm{core}$, and compute the central pressure
  via the equation of state, 
  $p_\mathrm{core} = p(\rho_\mathrm{core},T_\mathrm{core},X_\mathrm{core})$.

\item We then construct our model by integrating outward from the
  center and iterate over the central density and transition density
  until the mass of the core, $M_\mathrm{WD}$ and the mass of the
  helium envelope, $M_\mathrm{He}$, are the desired values.

  To be more precise, given $\rho_\mathrm{core}$ and
  $\rho_\mathrm{He}$, we do the following:
  \begin{itemize}

  \item Specify the composition for zone $i$:
        \begin{equation}
        X_i = 
        \left \{ \begin{array}{ll}
             X_\mathrm{core}  & \mathrm{if~} \rho_{i-1} > \rho_\mathrm{He} \\
             X_\mathrm{core} + \frac{1}{2} (X_\mathrm{He} - X_\mathrm{core}) 
                 \left [ 1 + \tanh\left(\frac{x_i - x_\mathrm{He} - 4\delta}{\delta} \right ) \right ] &
                 \mathrm{if~} \frac{1}{2} \left [ 1 + \tanh\left(\frac{x_i - x_\mathrm{He} - 4\delta}{\delta}\right ) \right ] < 0.999 \\
             X_\mathrm{He} & \mathrm{otherwise}
        \end{array} \right .
        \end{equation}
        here, $\delta$ is the width of the transition layer and
        $x_\mathrm{He} = x(\rho_\mathrm{He})$ is the coordinate
        corresponding to $\rho_\mathrm{He}$.  It is important that
        $\delta$ be resolved on our grid. 

  \item Specify the temperature
        \begin{equation}
        T_i = 
        \left \{ \begin{array}{ll}
             T_\mathrm{core}  & \mathrm{if~} \rho_{i-1} > \rho_\mathrm{He} \\
             T_\mathrm{core} + \frac{1}{2} (T_\mathrm{base} - T_\mathrm{core}) 
                 \left [ 1 + \tanh\left(\frac{x_i - x_\mathrm{He} - 4\delta}{\delta} \right ) \right ] &
                 \mathrm{if~} \frac{1}{2} \left [ 1 + \tanh\left(\frac{x_i - x_\mathrm{He} - 4\delta}{\delta}\right ) \right ] < 0.999 \\
             \max\{T(s_\mathrm{He},\rho_i,X_i), T_\mathrm{cutoff}\} & \mathrm{otherwise}
        \end{array} \right .
        \end{equation}
        here, $T_\mathrm{base}$ is the desired temperature at the base
        of the He layer, and $s_\mathrm{He}$ is the specific entropy
        of the base, $s_\mathrm{He} = s(\rho_\mathrm{base}, T_\mathrm{base},
        X_\mathrm{base})$.  Finally, $T_\mathrm{cutoff}$ is the lowest
        temperature allowed in the outer envelope.

        In both of these profiles, a tanh profile was used at the base 
        of the layer.  The true functional form of the transition is 
        not known.  Adopting a tanh gives it a smoothness that is desirable
        for hydrodynamics codes while still keeping the transition narrow.

  \item Compute the pressure using the equation of hydrostatic equilibrium (HSE).
        We difference the HSE equation as:
      \begin{equation}
      p^\mathrm{HSE}_i = p_{i-1} + \frac{1}{2} \Delta r (\rho_i + \rho_{i-1}) g_{i-1/2}
      \end{equation}
      Here $g_{i-1/2}$ is the gravitational acceleration at the lower
      edge of the zone, computed as $g_{i-1/2} =
      -GM_\mathrm{encl}/x_{i-1/2}^2$, where $M_\mathrm{encl}$ is the
      mass integrated up to that edge.  Given our guess for the
      density in the zone, $\rho_i$, the equation of state will return
      a pressure $p^\mathrm{EOS}_i = p(\rho_i,T_i,X_i)$.  We define a
      function, $F$, as $F = p^\mathrm{HSE}_i - p^\mathrm{EOS}_i$ and
      use a Newton-Raphson iteration to find the $\rho_i$ needed to
      zero $F$.  In the case where we are constraining the profile to
      be isentropic (with constant entropy $s_\mathrm{He}$) we begin
      with a guess for $T_i$, and zero an additional function, $G =
      s_\mathrm{He} - s^\mathrm{EOS}_i$, where $s^\mathrm{EOS}_i =
      s(\rho_i,T_i,X_i)$.  In this case, zeroing both $F$ and $G$
      yield $\rho_i$ and $T_i$.
      
  \item After all zones are computed, compute $M_\mathrm{WD}$ by
    integrating up all the carbon (and oxygen if present), and
    $M_\mathrm{He}$ by integrating up all the helium.  We then correct
    $\rho_\mathrm{core}$ and $\rho_\mathrm{He}$ using a secant method,
    recompute $p_\mathrm{core}$, and iterate the above procedure until
    the model is converged.

  \end{itemize}
\end{itemize}

\begin{figure}
\centering
\includegraphics[scale=0.7]{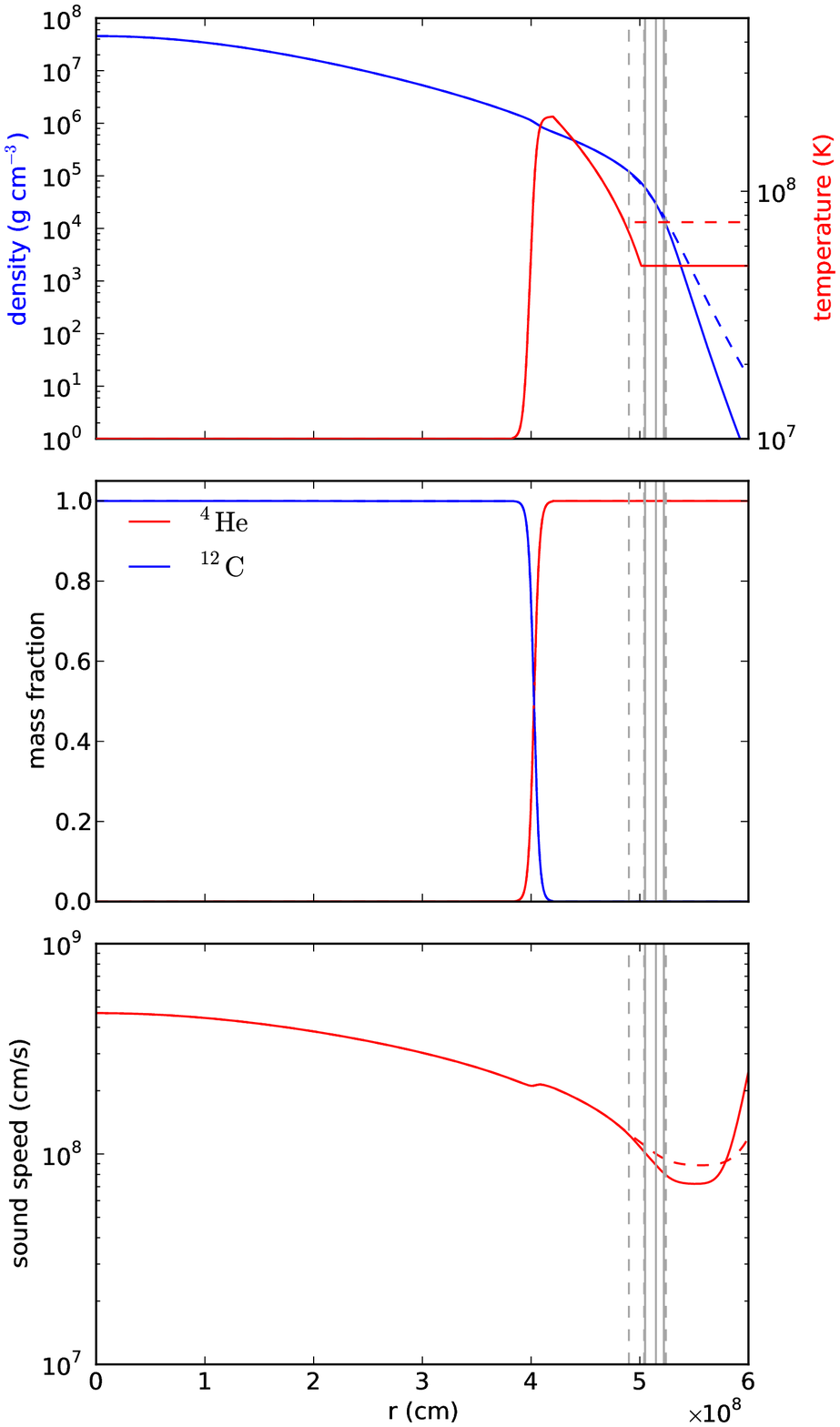}
\caption{\label{fig:initial_model} The initial model.  The solid lines
  are for the cooler cutoff model ($T_\mathrm{cutoff} = 5\times
  10^7$~K) and the dashed lines are for the hotter cutoff model
  ($T_\mathrm{cutoff} = 7.5\times 10^7$~K).  The vertical lines
  indicate the start of the sponge (leftmost), the anelastic cutoff
  (center), and the base cutoff density (rightmost), again with solid
  for the cooler cutoff model and dashed for the hotter cutoff model.}
\end{figure}

For the current simulations, we use two slightly different models.  For
both, we set $X_\mathrm{core}$ to be pure carbon and $X_\mathrm{He}$
to be pure helium.  (The reason we leave oxygen out of the white dwarf
for these initial models is to allow us to easily use the oxygen
generated by the reactions as a tracer of the nucleosynthesis).  We
pick $M_\mathrm{WD} = 1~\msolar$ (here, $M_\mathrm{WD}$ is the
integrated mass of the carbon only), $M_\mathrm{He} = 0.05~\msolar$,
$T_\mathrm{core} = 10^7~\mathrm{K}$, $T_\mathrm{base} = 2\times
10^8~\mathrm{K}$, and $\delta = 5\times 10^6~\mathrm{cm}$.  This value
of $\delta$ ensures that the composition gradients are somewhat
smoothed.  The only difference between the models is the choice of
cutoff temperature above the convective zone.  Our cool model has
$T_\mathrm{cutoff} = 5\times 10^7~\mathrm{K}$, and our hotter model
has $T_\mathrm{cutoff} = 7.5\times 10^7~\mathrm{K}$.
Figure~\ref{fig:initial_model} shows the profile for both models.  The
cool model is indicated by the solid lines and the hot model is the
dashed lines.  When creating these models, we pick a zone width, $\Delta r$,
to be $1/5^\mathrm{th}$ of the Cartesian zone width, $\Delta x$.

It is interesting to look at the behavior of the sound speed in the
initial model, shown in the lower panel of
Figure~\ref{fig:initial_model}.  For very low Mach number flows, the
time step for a compressible algorithm (assuming uniform zoning) will
be set where the sound speed is highest---the center of the star.  For
the low Mach number algorithm, it will be set where the velocities are
the highest---presumably in the convectively unstable region.  But the
highest Mach number in our simulations is not necessarily there as
well---it is likely to be at the very edge of the star, where the
velocity will rise due to the density gradient, but the sound speed
is small relative to the sound speed at the core.  For this reason, we may realize a
moderately large Mach number just outside the star ($\sim 0.2$--$0.3$),
but still be able to take a time step an order-of-magnitude larger than
a compressible code, because the peak Mach number is not where the
sound speed peaks.
We note that the upturn in sound speed at the largest radii shown in 
Figure~\ref{fig:initial_model} 
is not mapped into our computational domain because of the use of
a low density cutoff.  This upturn arises because of the
dominance of the radiation pressure in the general equation of state,
a behavior that is unphysical outside the star.

\subsection{Simulation Parameters}

We use the same set of cutoff densities and sponging techniques described 
in \citet{wdconvect,wdTurb}.  These parameters are designed to eliminate large
velocities that arise due to the steep density gradient at the edge of
the star, outside of our region of interest.  

\begin{figure}
\centering
\includegraphics[scale=0.85]{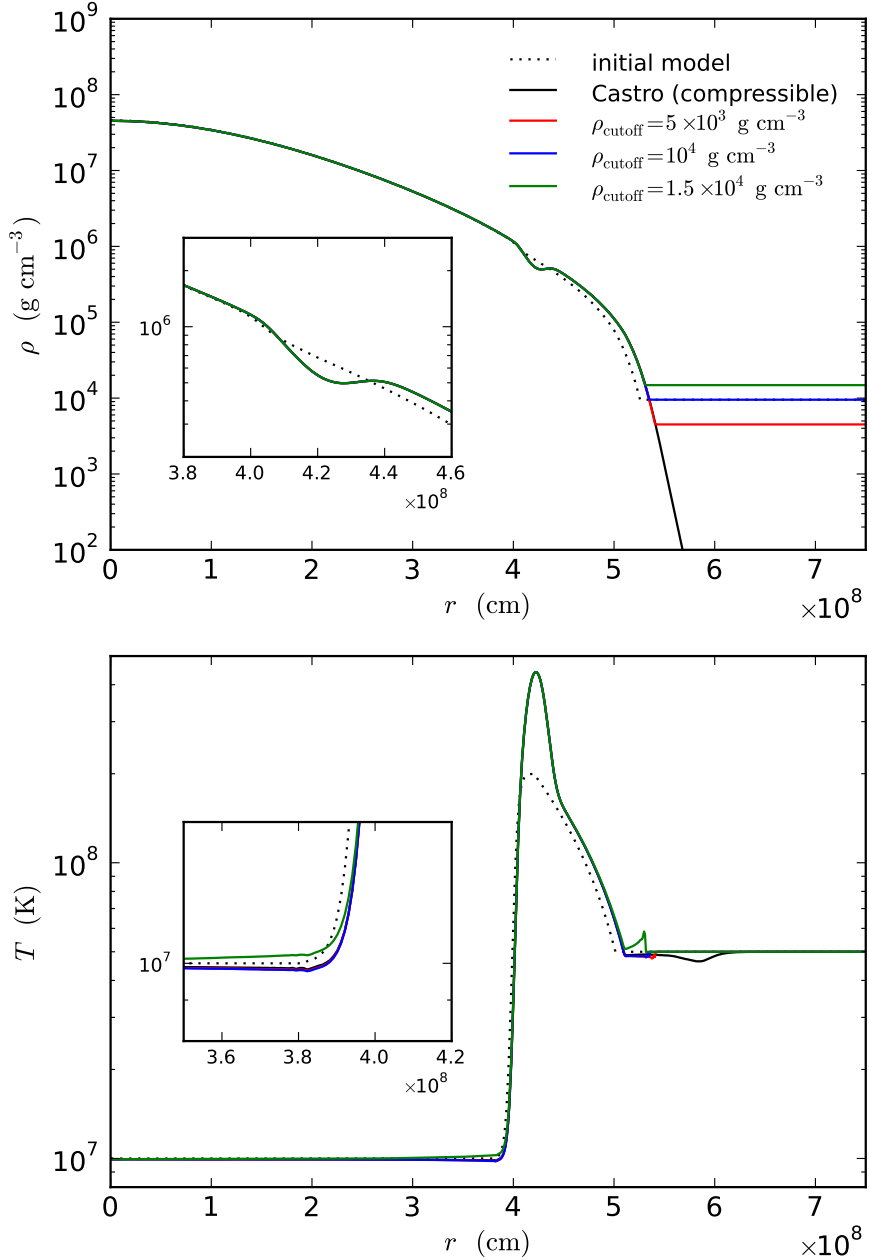}
\caption{\label{fig:base_expand} A test of the base state expansion
  algorithm.  The initial model is shown as the dotted line (for the
  choice $\rhocutoff = 10^{4}~\gcc$).  The compressible solution by
  the CASTRO code is shown as the black line, and three MAESTRO
  solutions are shown, with varying $\rhocutoff$.  We
  see very good agreement across the board between MAESTRO and CASTRO,
  with the two lowest cutoff densities showing the best match.}
\end{figure}

A low density cutoff, $\rhocutoff$, is the minimum density used in the
initial model---outside of that radius in the star, $r_\mathrm{cutoff}
= r(\rhocutoff)$, we hold the density constant.
The material at $r > r_\mathrm{cutoff}$ does not contribute to the computation of the
gravitational acceleration.  To determine a reasonable value of
$\rhocutoff$, we perform a one-dimensional base state expansion test,
comparing the results to the compressible code CASTRO~\citep{castro}.
Figure~\ref{fig:base_expand} shows the results of placing our initial
model onto a one-dimensional spherical grid and heating it for 25~s with a heating
term:
\begin{equation}
H = A_\mathrm{heat} X(^{4}\mathrm{He}) e^{-(r - r_\mathrm{heat})^2/\sigma_\mathrm{heat}^2}
\end{equation}
We choose $A_\mathrm{heat} = 10^{15}~\mathrm{erg~g^{-1}~s^{-1}}$,
$r_\mathrm{heat} = 4.2\times 10^8~\mathrm{cm}$, and
$\sigma_\mathrm{heat} = 10^7~\mathrm{cm}$.  These values were 
chosen to concentrate the energy release at the base of the He layer
and to ensure that we see a significant response to the hydrostatic
structure.   The base state expansion
algorithm used in MAESTRO is described in detail
in~\citet{multilevel}.  We run with three different choices of
$\rhocutoff$: $5\times10^3$, $10^4$, and $1.5\times10^4$ g cm$^{-3}$.
The figure shows excellent agreement between the fully compressible
(CASTRO) results and MAESTRO's base state expansion.  We note that
this is a rather severe test, and the amount of expansion seen here is
greater than what we expect in our three-dimensional simulations.
There is a slight departure from the compressible solution at the base
of the helium layer in the temperature field for the largest choice of
$\rhocutoff$.  Based on these results, we choose $\rhocutoff =
10^4~\gcc$ for the simulations presented here.  As in
\citet{wdconvect,wdTurb} we have a cutoff to the buoyancy term in the
momentum equation which we set as $2\rhocutoff$.  It is important to
note that this one-dimensional test does not have any transport of the
energy, it simply expands the hydrostatic structure in response to the
heating.  As a result, the large increase in temperature seen here
will not appear in the actual three-dimensional simulation (since
convection would redistribute the heat).

Outside of the star, we want to damp away large velocities, as this region
is not really part of the simulation space and we do not want velocities
here to build up and influence our time step.  In MAESTRO, this is 
accomplished through the use of a sponge, which appears as a source
in the velocity equation.
The sponge has the same functional form as presented in \citet{paper4}
and as refined in \citet{wdconvect,wdTurb}.  In particular, we define a
density at which to center the sponge, $\rho_\mathrm{md}$, and a
multiplicative factor, $f_\mathrm{sp}$, to mark the start of the
sponge.  This means that the sponging turns on (gradually) once the
model drops below a density of $f_\mathrm{sp} \rho_\mathrm{md}$.  We
chose the sponge parameters to make the sponge turn on at the
top of the convective layer.  Finally, we modify the constraint
equation to act like the anelastic constraint to suppress high
velocities at the outer boundary of the star (see \citealt{ABNZ:III}).
This occurs once the density drops below $\rhoanelastic$.
Table~\ref{table:cutoffs} summarizes the cutoff parameters for the two
different initial models.  These parameters are illustrated in
Figure~\ref{fig:initial_model} as the vertical lines, again with the
solid corresponding to the cooler cutoff model and the dashed to the
hotter cutoff model.

\subsection{Grid Structure}

\begin{figure}
\centering
\includegraphics[scale=0.25]{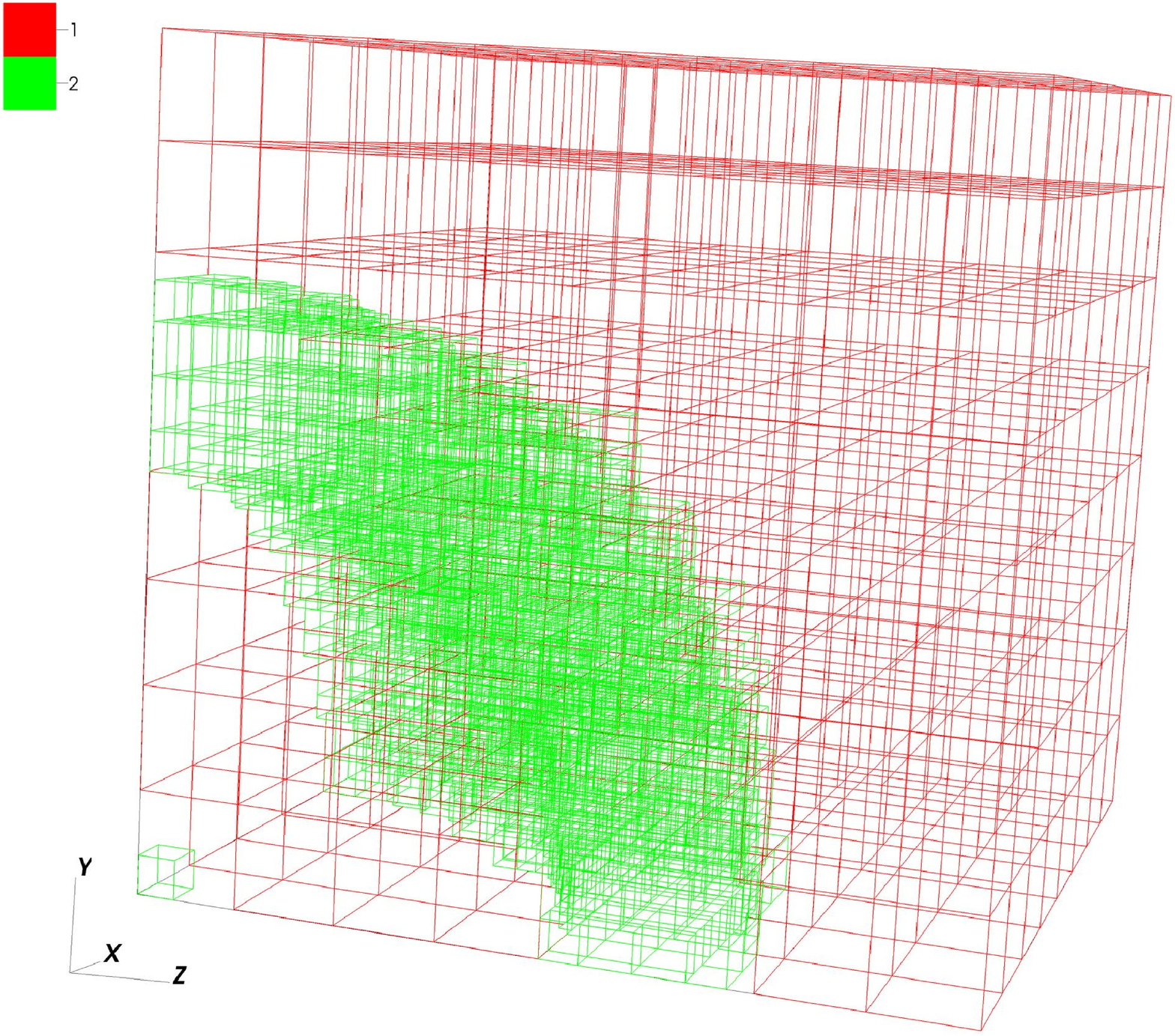}
\caption{\label{fig:amr} A representative 2-level grid used for the
simulations.}
\end{figure}

Adaptive mesh refinement is used, with the refinement tagging on zones
that have $X(^4\mathrm{He}) > 0.01$ and $\rho \ge
\rho_\mathrm{cutoff}$.  Additionally, we always refine the very center
of the star (the coordinate origin) due to the design of the averaging
algorithm from the Cartesian grid to the radial base state
\citep{multilevel}.  Simulations are run with one level of refinement
with a factor of two increase in resolution on top of the coarse
grid---Figure~\ref{fig:amr} shows a representative grid structure.
The work is parallelized by distributing grids to nodes that communicate
with each other using MPI and OpenMP to spawn threads within nodes to
perform floating point work on the data.
Unless otherwise noted, all simulations use a
base grid of $256^3$ with an additional level of refinement on the
helium layer.  The computational domain is a cube with a side of
$7.5\times 10^8~\mathrm{cm}$, with the center of the star placed at
the origin.  This gives a resolution of $\Delta x^\mathrm{fine} =
14.6~\mathrm{km}$ in the convective region.  The one-dimensional
radial grid for the base state uses a resolution $\Delta r = \Delta
x^\mathrm{fine}/5$, to improve the performance of the mapping between
the Cartesian and radial grids (see \citealt{multilevel}).  The
boundary conditions are reflecting on the symmetry faces of the domain
(lower $x$, $y$, and $z$) and outflow (zero-gradient) on the other
faces.

\subsection{Velocity Field Initialization}\label{sec:velocity init}

We define the initial velocity field to be a perturbation with similar form
to that described in~\citet{paper4}.  For a zone with coordinates
$(x,y,z)$, a set of Fourier modes is defined as:
\begin{equation}
C^{(x)}_{l,m,n} = \cos \left (\frac{2 \pi l x}{\sigma} + \phi^{(x)}_{l,m,n} \right ),
\quad
C^{(y)}_{l,m,n} = \cos \left (\frac{2 \pi m y}{\sigma} + \phi^{(y)}_{l,m,n} \right ),
\quad
C^{(z)}_{l,m,n} = \cos \left (\frac{2 \pi n z}{\sigma} + \phi^{(z)}_{l,m,n} \right ),
\end{equation}
\begin{equation}
S^{(x)}_{l,m,n} = \sin \left (\frac{2 \pi l x}{\sigma} + \phi^{(x)}_{l,m,n} \right ),
\quad
S^{(y)}_{l,m,n} = \sin \left (\frac{2 \pi m y}{\sigma} + \phi^{(y)}_{l,m,n} \right ),
\quad
S^{(z)}_{l,m,n} = \sin \left (\frac{2 \pi n z}{\sigma} + \phi^{(z)}_{l,m,n} \right ),
\end{equation}
where $\sigma$ is the perturbation scale and
$\phi^{\{x,y,z\}}_{l,m,n}$ are randomly generated phases that lie
between 0 and $2\pi$.  The velocity perturbation in the zone is then
set as:
\begin{mathletters}
\begin{eqnarray}
u' &=& \sum_{l=1}^3 \sum_{m=1}^3 \sum_{n=1}^3 \frac{1}{N_{l,m,n}}
   \left [ - \gamma_{l,m,n} m C^{(x)}_{l,m,n} S^{(y)}_{l,m,n} C^{(z)}_{l,m,n} 
           + \beta_{l,m,n}  n C^{(x)}_{l,m,n} C^{(y)}_{l,m,n} S^{(z)}_{l,m,n} \right ], \\
v' &=& \sum_{l=1}^3 \sum_{m=1}^3 \sum_{n=1}^3 \frac{1}{N_{l,m,n}}
   \left [\phantom{+}  \gamma_{l,m,n} l S^{(x)}_{l,m,n} C^{(y)}_{l,m,n} C^{(z)}_{l,m,n} 
           - \alpha_{l,m,n} n C^{(x)}_{l,m,n} C^{(y)}_{l,m,n} S^{(z)}_{l,m,n} \right ], \\
w' &=& \sum_{l=1}^3 \sum_{m=1}^3 \sum_{n=1}^3 \frac{1}{N_{l,m,n}}
   \left [ - \beta_{l,m,n}  l S^{(x)}_{l,m,n} C^{(y)}_{l,m,n} C^{(z)}_{l,m,n} 
           + \alpha_{l,m,n} m C^{(x)}_{l,m,n} S^{(y)}_{l,m,n} C^{(z)}_{l,m,n}  \right ],
\end{eqnarray}
\end{mathletters}
where the amplitudes $\alpha_{l,m,n}$, $\beta_{l,m,n}$, and
$\gamma_{l,m,n}$ are randomly generated to lie between $-1$ and 1, and
$N_{l,m,n} = \sqrt{l^2 + m^2 + n^2}$.

Finally, we confine the perturbation to lie in the convective region as:
\begin{mathletters}
\label{eq:velpert}
\begin{eqnarray}
u'' &=& \frac{A u' }{4} 
  \left [ 1 + \tanh \left ( \frac{r_\mathrm{pert}^\mathrm{outer} - r - d}{d} \right ) \right ]
  \left [ 1 + \tanh \left ( \frac{r - r_\mathrm{pert}^\mathrm{inner} - d}{d} \right ) \right ],\\
v'' &=& \frac{A v' }{4} 
  \left [ 1 + \tanh \left ( \frac{r_\mathrm{pert}^\mathrm{outer} - r - d}{d} \right ) \right ]
  \left [ 1 + \tanh \left ( \frac{r - r_\mathrm{pert}^\mathrm{inner} - d}{d} \right ) \right ],\\
w'' &=& \frac{A w' }{4} 
  \left [ 1 + \tanh \left ( \frac{r_\mathrm{pert}^\mathrm{outer} - r - d}{d} \right ) \right ]
  \left [ 1 + \tanh \left ( \frac{r - r_\mathrm{pert}^\mathrm{inner} - d}{d} \right ) \right ].
\end{eqnarray}
\end{mathletters}
Here, the first $\tanh$ factor cuts the perturbation off at the outer
edge of the star and the second $\tanh$ factor cuts the perturbation
off at the base of the convective boundary.  These transitions are
characterized by a thickness $d$.  $A$ is the amplitude of the
perturbation.  For all the simulations presented here, we choose $A =
10^5~\mathrm{cm~s^{-1}}$, $\sigma = 5\times 10^7~\mathrm{cm}$, and $d
= 10^7~\mathrm{cm}$.  The inner extent of the perturbation,
$r_\mathrm{pert}^\mathrm{inner}$ is set to be the radius where
$X(^4\mathrm{He}) > 0.9$ (moving from the center of the star to the
edge).  The outer extent is set as $r_\mathrm{pert}^\mathrm{outer} =
[r_\mathrm{pert}^\mathrm{inner} + r(f_\mathrm{sp}
  \rho_\mathrm{md})]/2$, where $r(f_\mathrm{sp} \rho_\mathrm{md})$ is
the radius where the sponge just begins to turn on.
This confines the velocity perturbation to the lower half of
the convectively unstable layer. 

\subsection{Hydrodynamic Integration Strategy}

The construction of advective fluxes on the faces of the computational
zones requires the prediction of values of the fluid state from the cell
centers to the faces at intermediate times.  
The method described in \citet{multilevel} predicts
the mass fractions, $X_k$, and the perturbational density,
$\rho^\prime$ (the difference between the zone's density and the
base state density at that radius), to the interface and algebraically
combines these with the
base state density to compute $(\rho X_k)$ on the edges.
Here, we instead predict the full density, $\rho$, and $X_k$ and define
the edge state as the product of these.  Numerical experiments show
that this variant is more robust at steep composition gradients (such as
the gradient at the base of our helium layer), in that unperturbed 
gradients become less smoothed.  This is similar to the
treatment of the density in our original implementation
\citep{ABRZ:I}.

\section{Results}\label{Sec:Results}

Unlike our previous simulations
of interior convection \citep{wdconvect,wdTurb}, in this problem convection
takes place on the surface of the star.  Here we will assess the 
applicability of our simulation methodology to modeling the convection 
in the helium layer on the surface of the white dwarf.
We discuss the general trends of the simulations and
explore the effect of various simulations parameters.
We defer a detailed discussion of the 
scientific implications and explorations of other white dwarf masses
to a later paper.

Our reference calculation is an octant simulation with the hot-$\Tcutoff$
initial model.  We will compare results from this reference calculation 
to a simulation with:
\begin{itemize}
\item disabled burning (Section \ref{sec:disabled burning}).
\item the full star rather than an octant (Section \ref{sec:full star}).
\item twice the spatial resolution (Section \ref{sec:resolution sensitivity}).
\item the cool-$\Tcutoff$ initial model rather than the 
      hot-$\Tcutoff$ model (Section \ref{sec:cool model}).
\item a modified treatment of the region beyond the convective layer, 
      as controlled by the cutoff densities and sponging 
      parameters (Section \ref{sec:sponge}).
\end{itemize}

\subsection{General Trends}

We begin by looking at the qualitative behavior of the convection
for our reference calculation.  Each time step we store the location 
of the hottest zone in the
convective region (only considering cells with $\rho > f_\mathrm{sp}
\rho_\mathrm{md}$).  We also store the maximum Mach number in the
entire computational domain.

\begin{figure}
\centering
\includegraphics[scale=0.3]{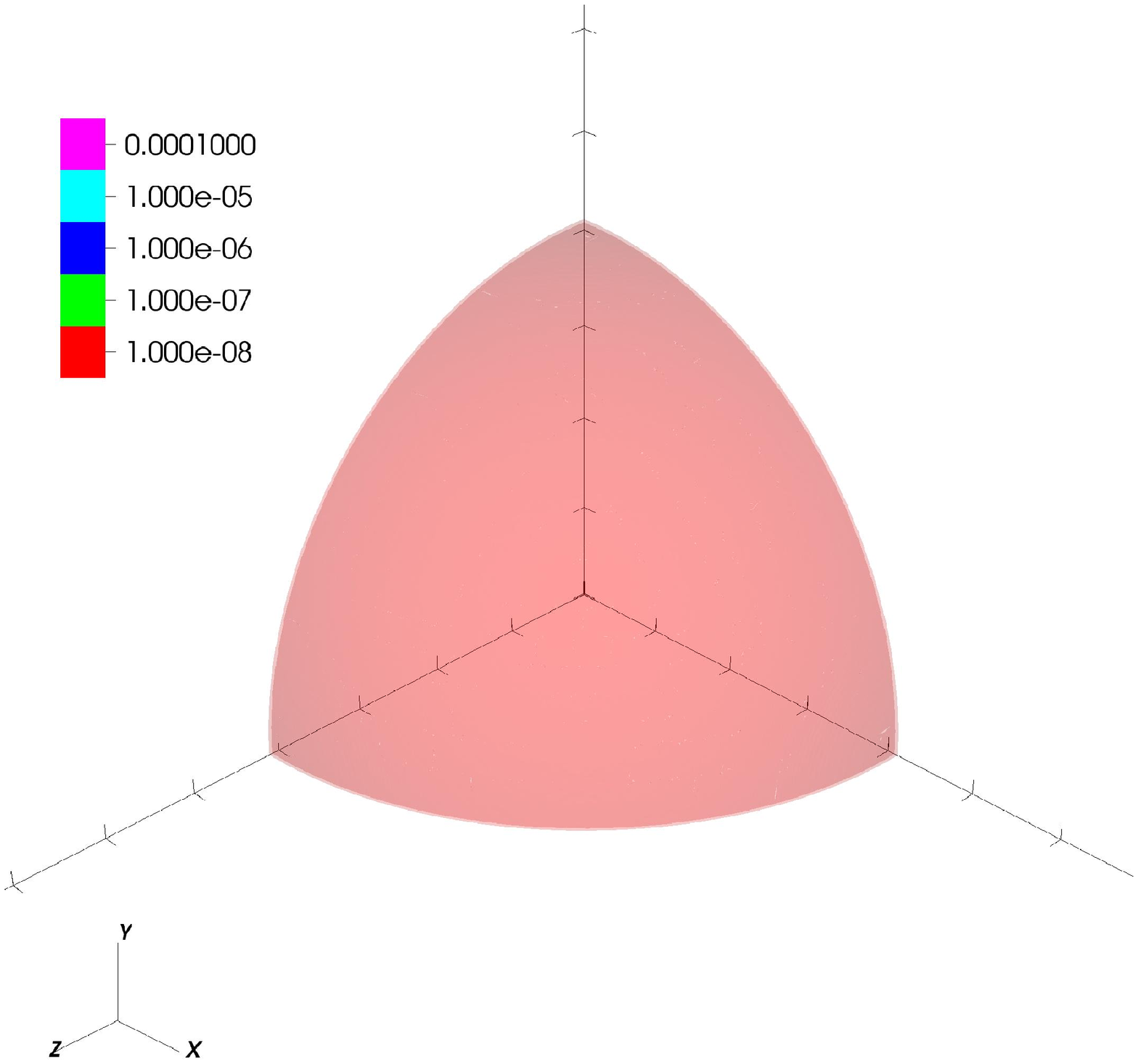}\hspace{1em}
\includegraphics[scale=0.3]{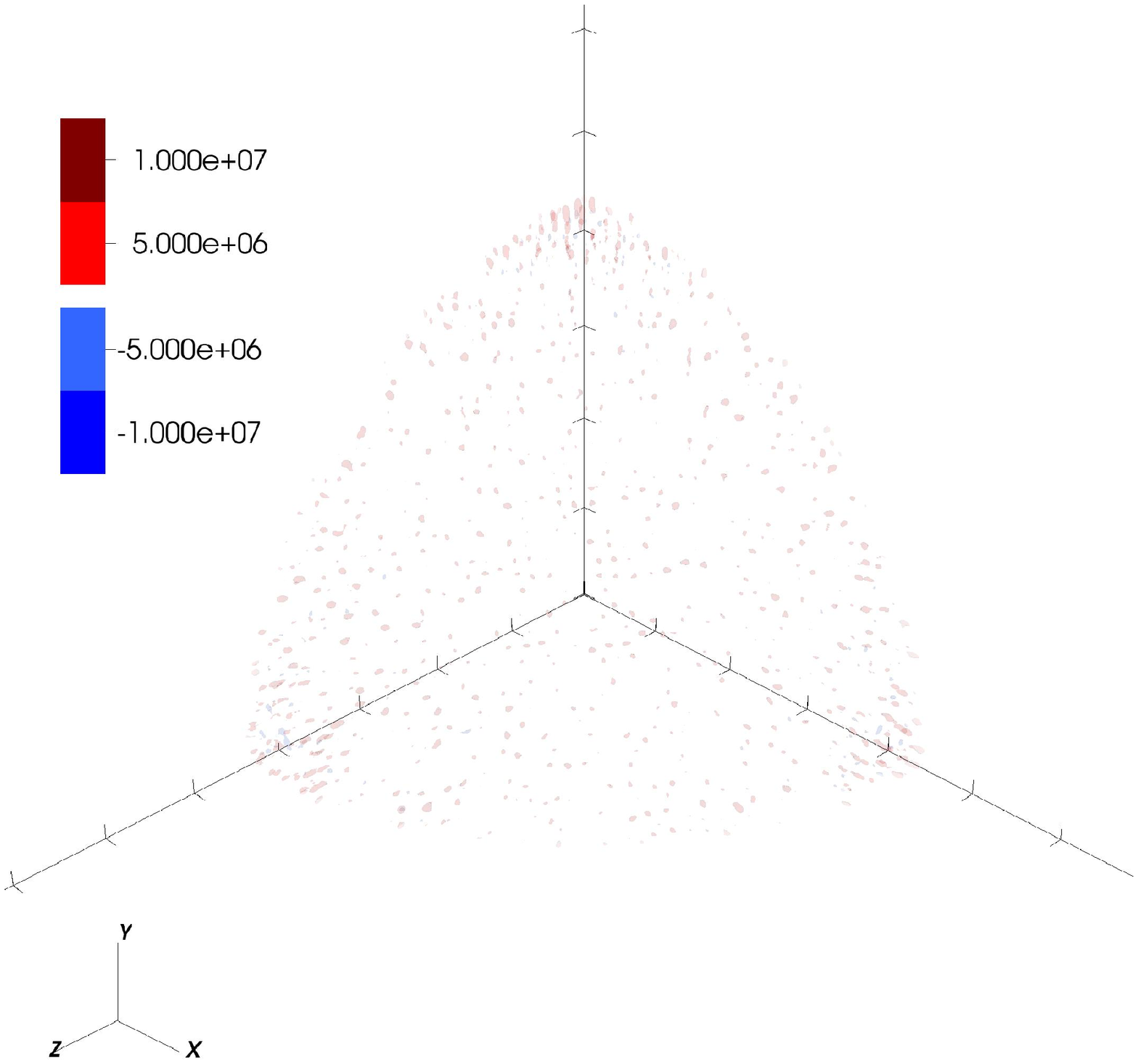} \\
\includegraphics[scale=0.3]{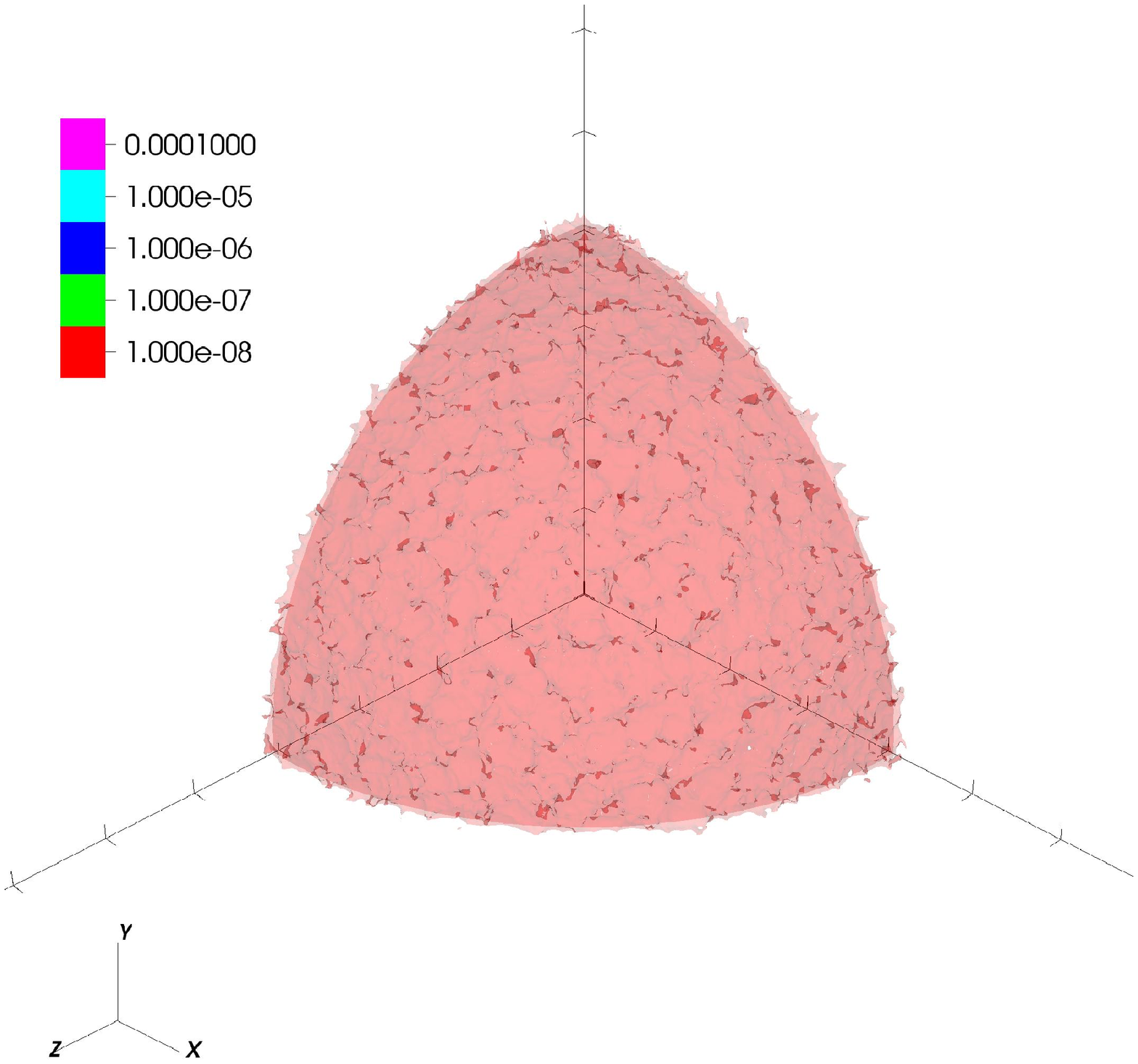}\hspace{1em}
\includegraphics[scale=0.3]{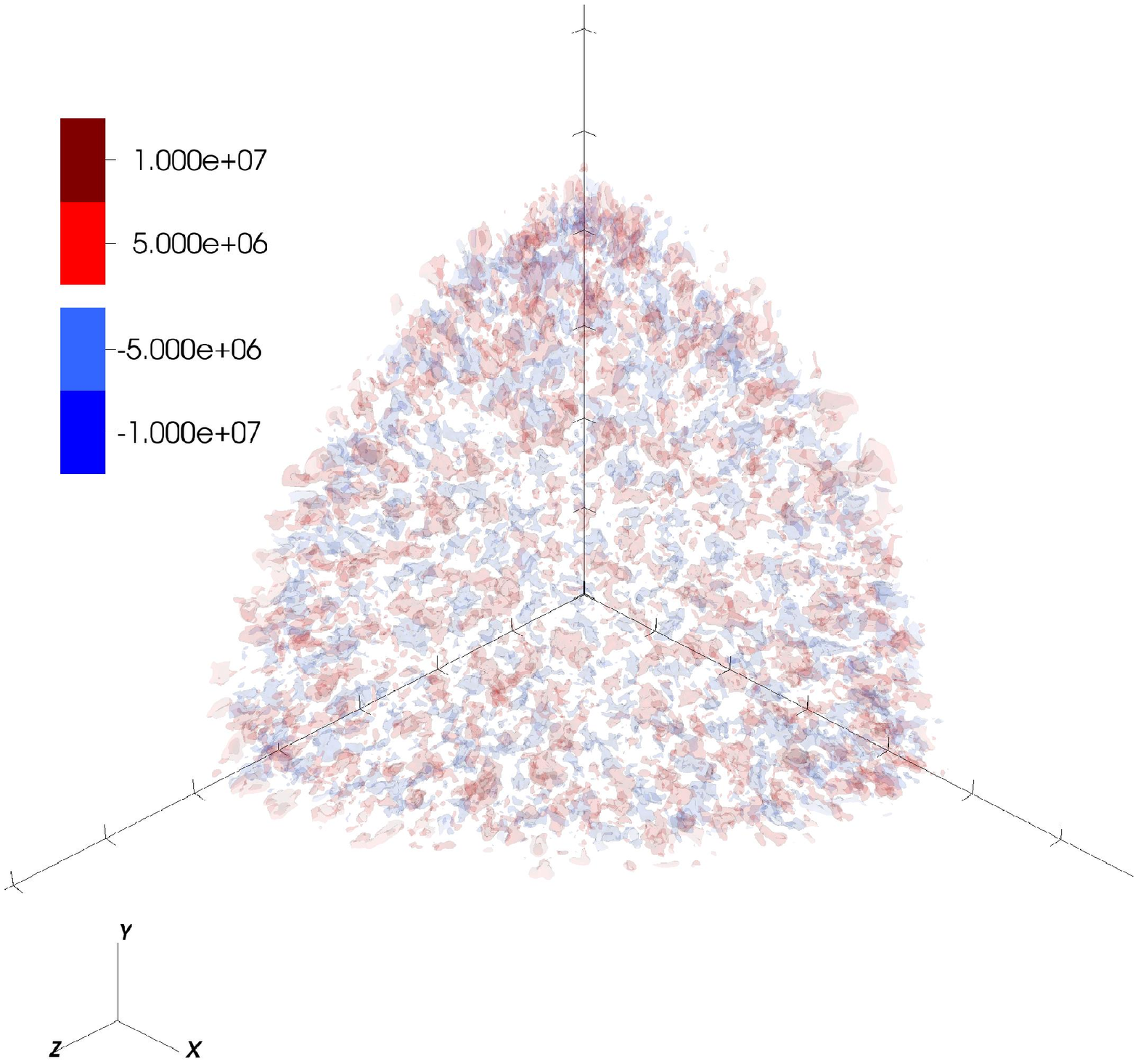} \\
\includegraphics[scale=0.3]{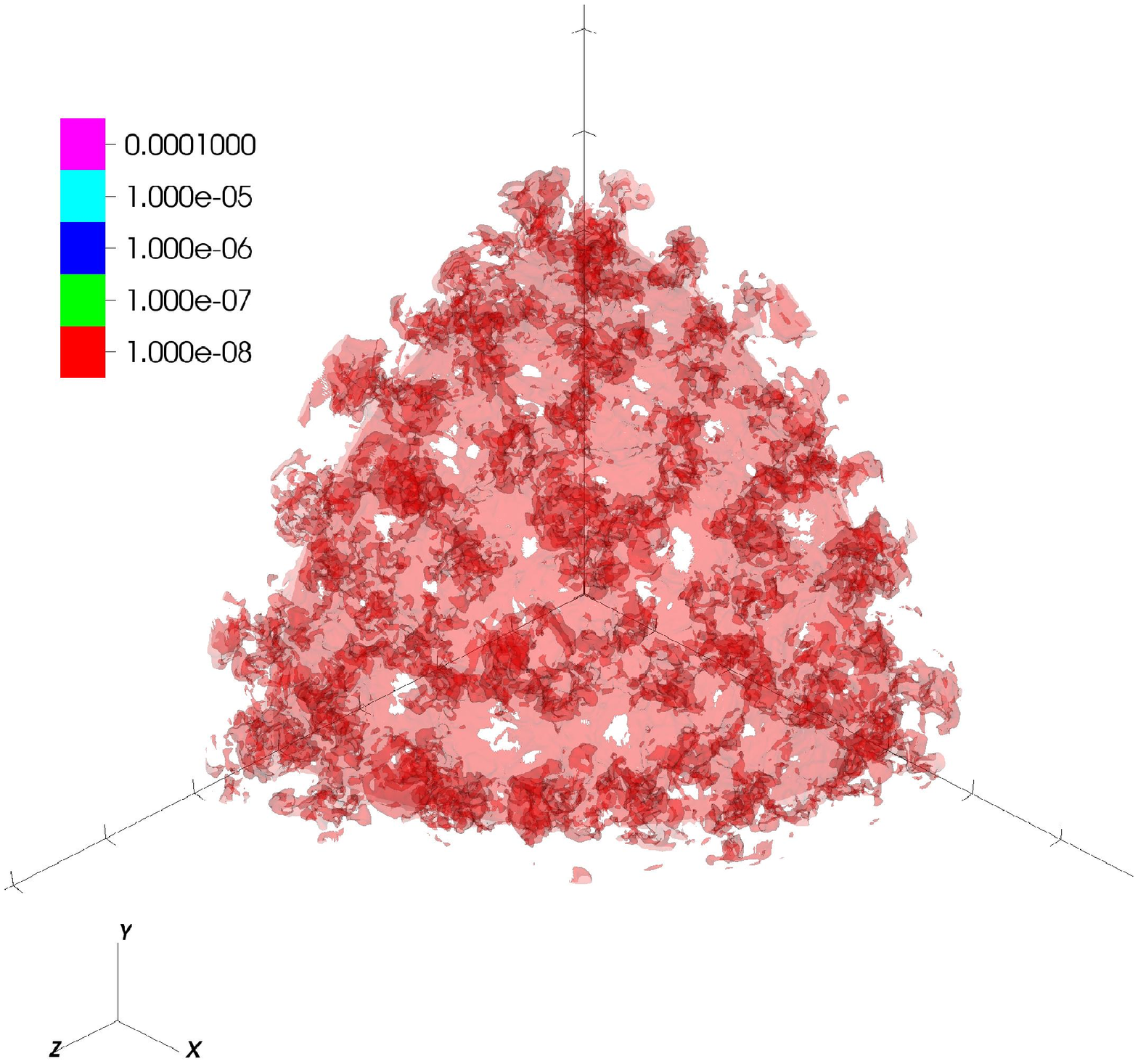}\hspace{1em}
\includegraphics[scale=0.3]{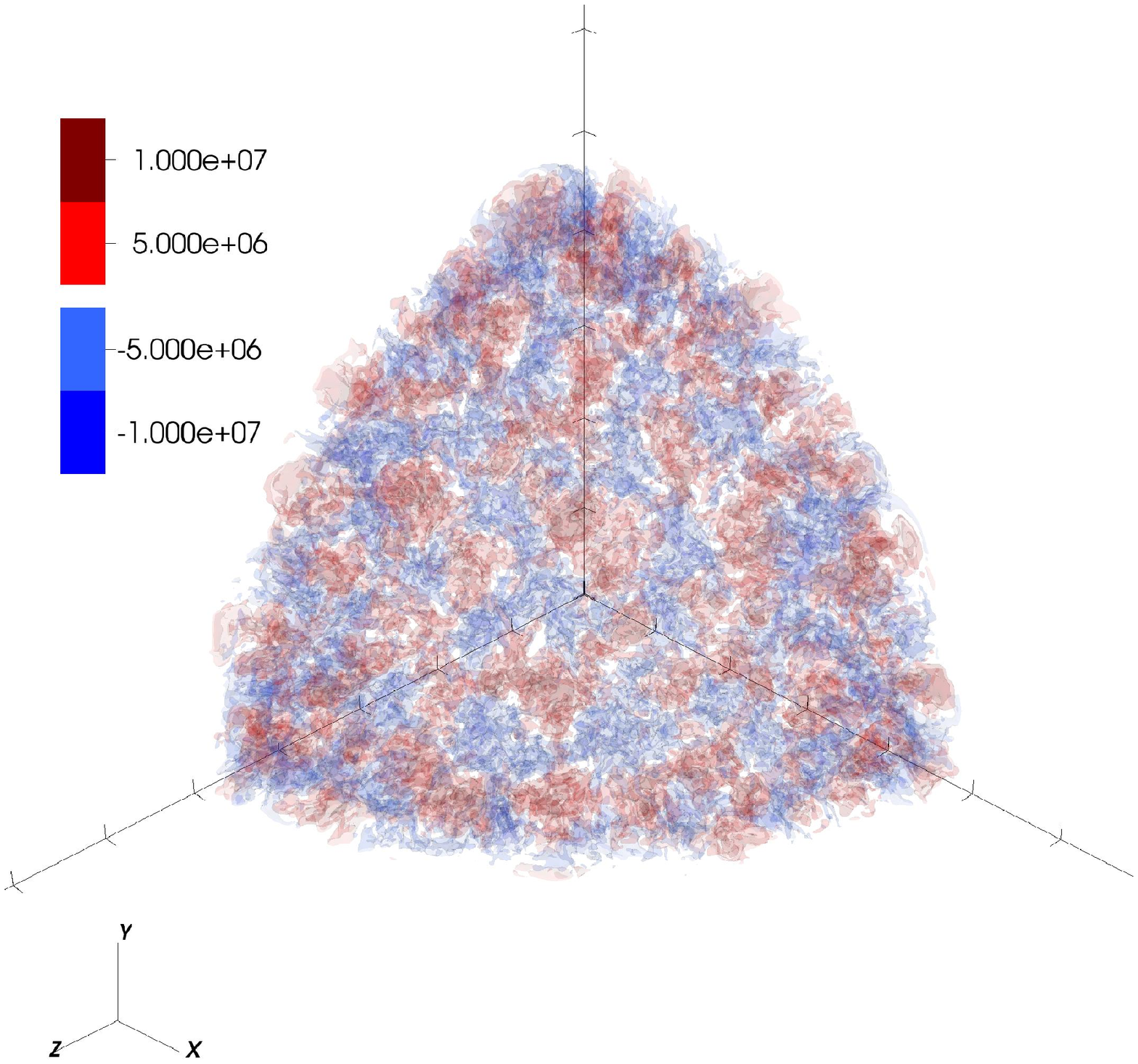} 
\caption{\label{fig:hot_baserun_seq} Time sequence of the
  $^{16}\mathrm{O}$ abundance (left) and radial velocity (right) for
  the hot-$\Tcutoff$ reference simulation shown at 20, 40, 60, 80, and
  100~s.  The tick marks are spaced $10^8$~cm apart.  The
  $^{16}\mathrm{O}$ sequence highlights the extent of the convective
  region well---by 80~s, there is a well-defined upper surface to the
  convective zone.  The radial velocity (CGS units); red=outflow,
  blue=inflow) shows a characteristic ``granulation'' pattern, making
  the distinct convective cells stand out.}
\end{figure}

Figure~\ref{fig:hot_baserun_seq} shows a time sequence for the 
reference calculation, with both the $^{16}\mathrm{O}$ abundance and radial
velocity (red = outflow; blue = inflow) visualized.  The development
of the convection is clearly seen.  By 80~s, the $^{16}\mathrm{O}$
synthesized at the base of the layer is distributed throughout the
entire convective zone, and a clear top to the shell is seen.  The
radial velocity shows that the convection is divided into cells, with
outflow regions surrounded by inflowing regions.  Looking at the 80~s plot, 
there are approximately 6 such cells along the edge of the domain.  The
circumference of this edge is $\sim (\pi /2) 5\times 10^8~\mathrm{cm}$
or $7.9\times 10^8~\mathrm{cm}$.  This tells us that a convective cell
has a diameter of approximately $1.3\times 10^8~\mathrm{cm}$.  This is
very close to the thickness of the convective layer ($\sim
10^8~\mathrm{cm}$ as seen in Figure~\ref{fig:initial_model}).
We note that the thickness of the convective layer is a function of the 
initial model, and we will explore other initial models in subsequent papers.

\begin{figure}
\centering
\includegraphics[scale=0.6]{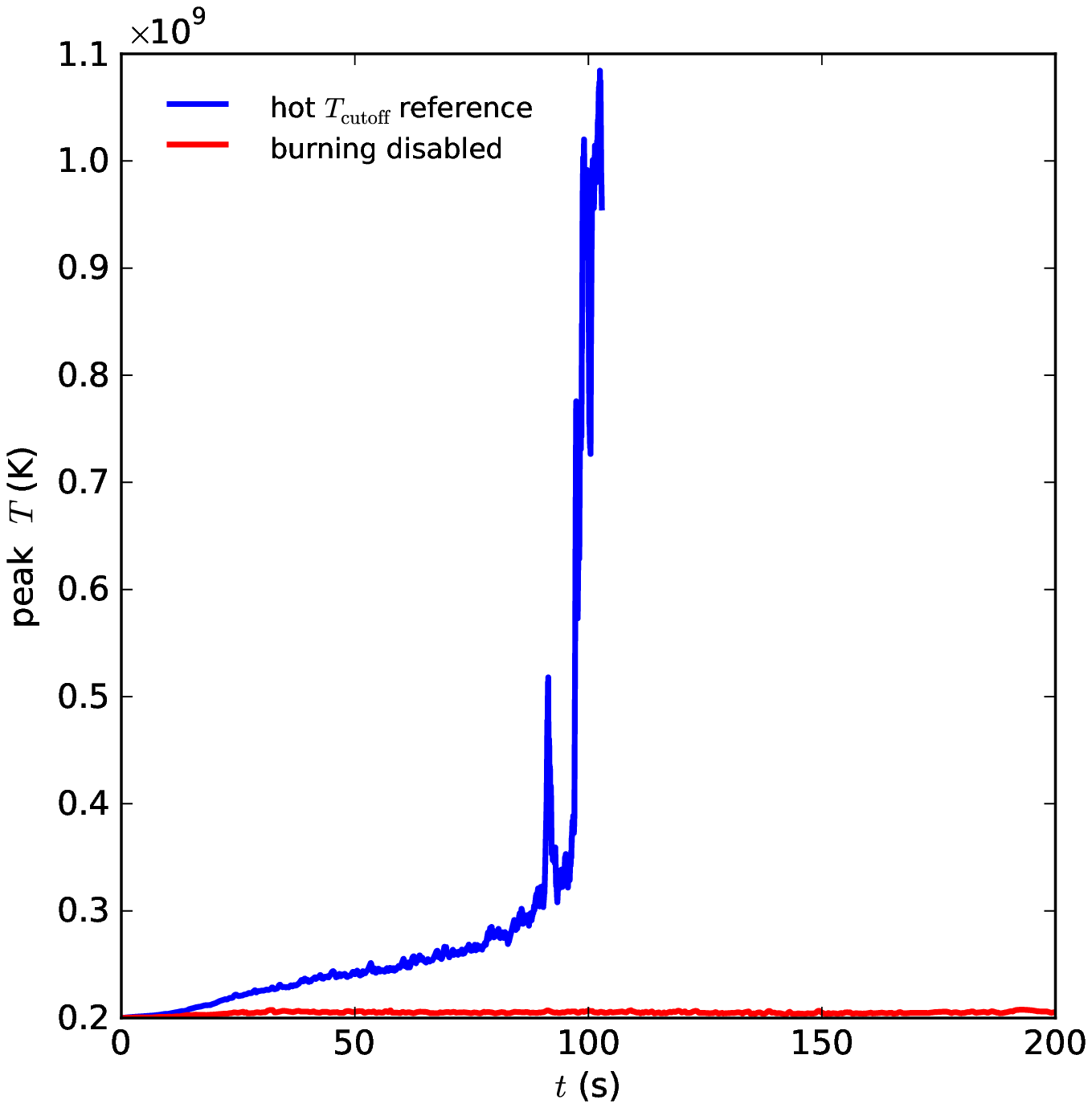} \\
\includegraphics[scale=0.6]{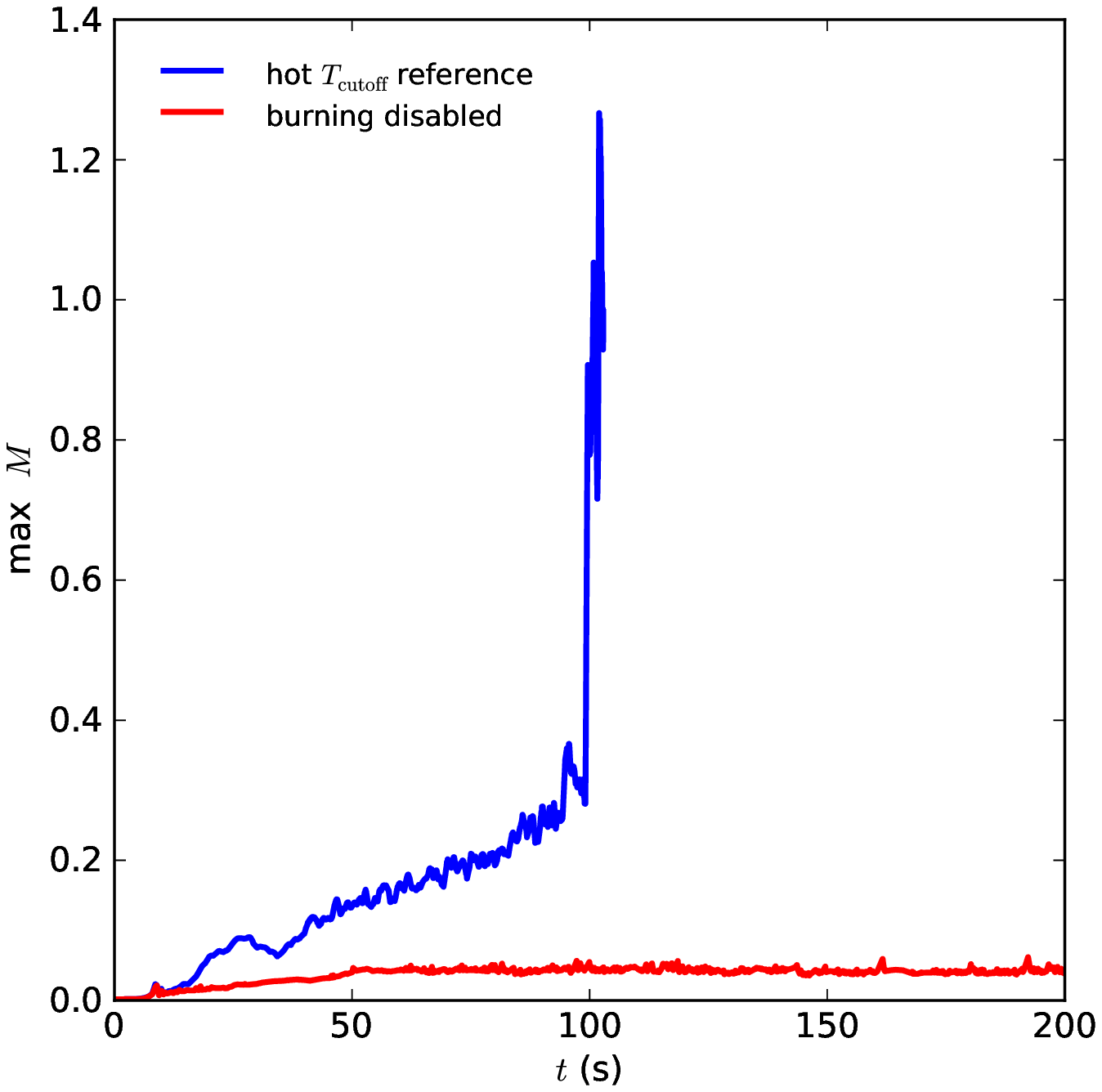} 
\caption{\label{fig:baserun} Peak temperature and Mach number as a
  function of time for our reference calculation (hot-$\Tcutoff$)
  and the same calculation with burning disabled.}
\end{figure}

Figure~\ref{fig:baserun} shows the peak temperature and peak Mach
number as a function of time for this run.  It is
interesting to see that there are a number of failed ignitions
toward the end---our suspicion is that the buoyant
hotspots rise and cool via expansion faster than the rate at which
energy is injected through reactions.  This is supported by the relic
plumes of $^{16}\mathrm{O}$ seen throughout 
Figure \ref{fig:hot_baserun_seq}.  A caveat 
to this behavior is that our network only continues to the production
of $^{16}\mathrm{O}$.  A more extensive reaction network might release
enough energy for the first hot spot to fully ignite. 

The Mach number panel in Figure~\ref{fig:baserun} shows that the Mach
number stays below $0.3$ for the bulk of the simulation.  
It is important to note that the efficiency metric for the low Mach
number algorithm, the increase in time step over a compressible code,
is not simply $1/M$ here.  For a compressible simulation, the higher
sound speed at the center of the star dominates the fluid
velocities realized at the edge of the star.  For these
calculations, we evolve to $100~s$ in $\sim 2100$ time steps, giving an
average $\Delta t_\mathrm{lowMach} \sim 0.05$.  The peak sound speed
in these models is $c_s = 4.7\times 10^8~\mathrm{cm~s^{-1}}$.  Taking
the coarse grid spacing, $\Delta x = 2.9\times 10^6~\mathrm{cm}$, the
compressible time step for these models would be $\Delta t = \Delta x /
c_s = 6.2\times 10^{-3}~\mathrm{s}$---a factor of 8 smaller.
Note that this doesn't factor any
of the fluid motions themselves into the compressible time step, which
would only further reduce it.   We also note that the early evolution
begins with a much smaller Mach number, so the efficiency is greatest
earlier in the evolution.  Finally, once a hotspot
ignites, the Mach number rises beyond the range of validity of the low Mach
number approximation, and we would need to transition this problem 
to a compressible code to continue.  Initial work on transitioning
MAESTRO calculations to the compressible code CASTRO
is shown in \cite{scidac-scaling}.

\begin{figure}
\centering
\includegraphics[width=6in]{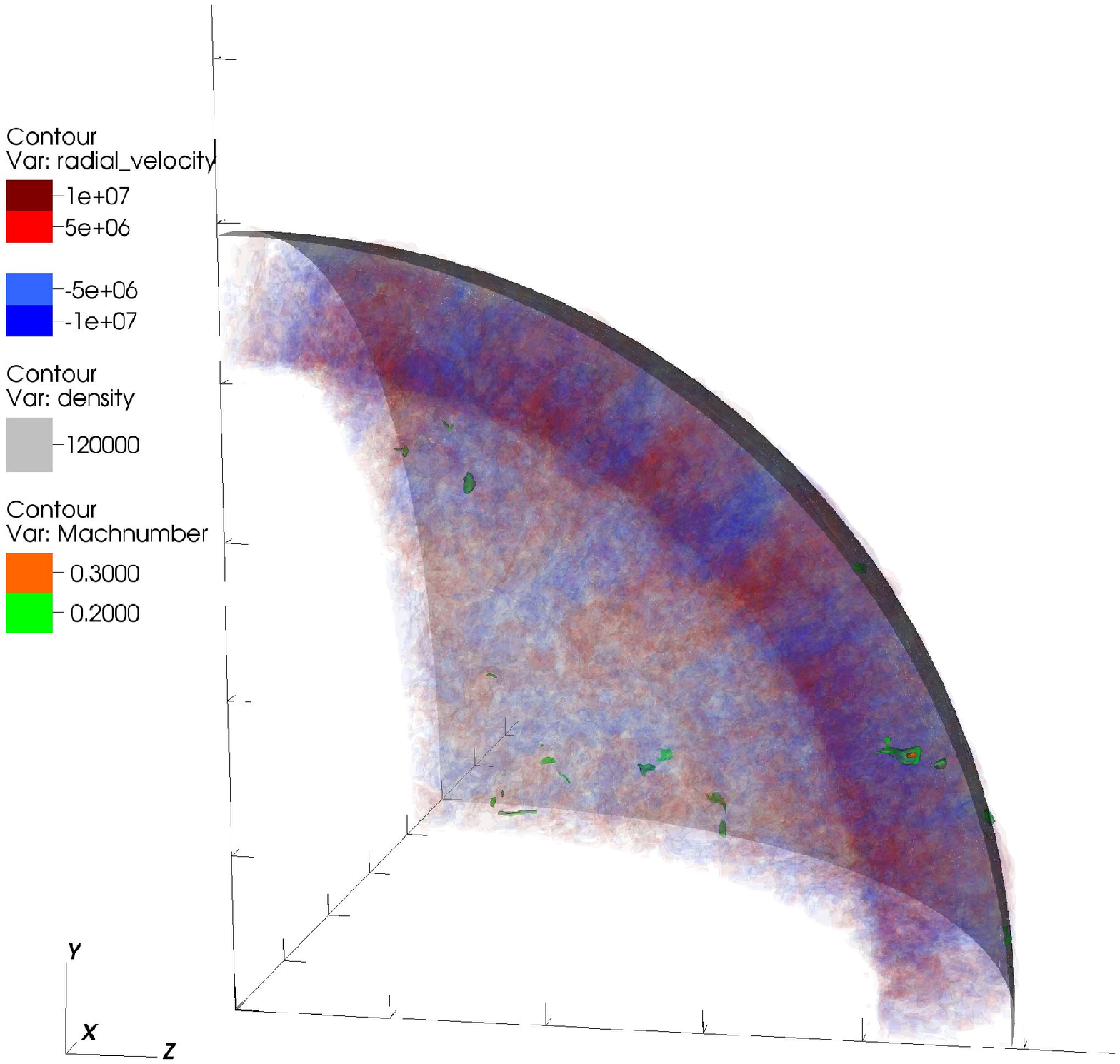}
\caption{\label{fig:hot_baserun_side} Side view of the convective
  field of the hot-$\Tcutoff$ reference run at 95~s.  All contours are
  in CGS units.  The tick marks are $10^8$~cm apart.  The grey density
  contour marks the surface where the sponging just begins to turn
  on---we see that the convection is entirely below this surface.  The
  orange and green contours show the locations where the Mach number
  is highest.}
\end{figure}

Figure~\ref{fig:hot_baserun_side} shows a side view of the convective
layer for the reference calculation.  We see that the convective layer
is well bounded.  The grey contour marks the density at which the
sponge just begins to turn on---as designed, this is at the upper
boundary of the convection.  The orange and green contours show the
regions where the Mach number is highest.  We see that these regions
are very small points, scattered throughout the layer, mostly near the
top of the convective zone.

\begin{figure}
\centering
\includegraphics[width=5in]{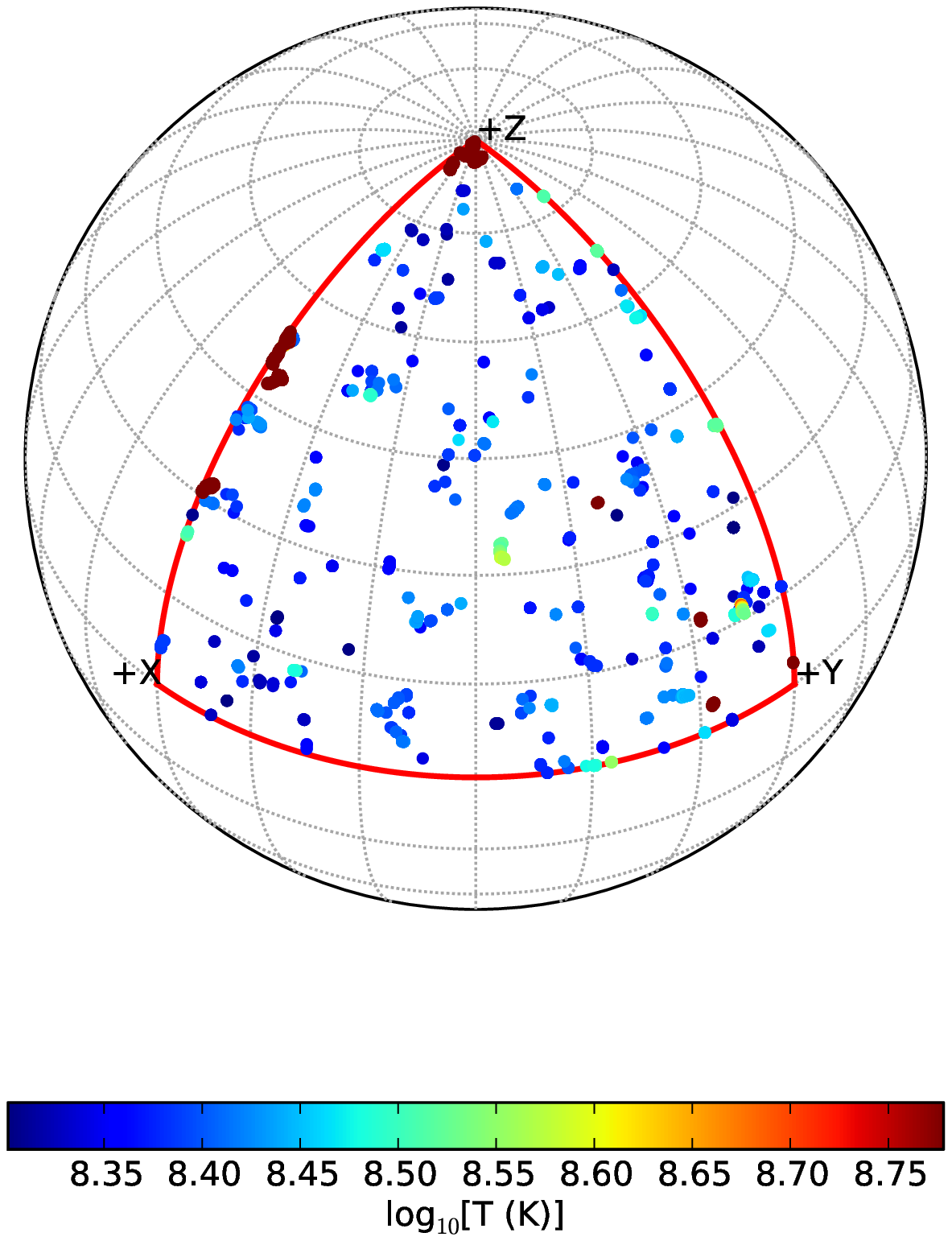}
\caption{\label{fig:baserun_hotspot} Location of the hottest point at
  each time step in the reference simulation, colored according to
  temperature.  The red boundary shows the bounds of the octant
  modeled.}
\end{figure}

Figure~\ref{fig:baserun_hotspot} shows the location of the hottest
point in the model (the hotspot) at each time step, colored by the
logarithm of the temperature.  We note that the time step changes
over the course of the simulation (generally speaking, smaller time steps
toward the end of the simulation).  The points where ignition
developed (seen, for example, as the $^{16}\mathrm{O}$ plumes in
the final time snapshot in
Figure~\ref{fig:hot_baserun_seq}) are shown in deep red.  We notice
that there seems to be a slight bias of ignition along the edges of the octant,
where the flow can be confined by the geometry.  We remind the reader
that all simulations are done in Cartesian coordinates---the meridians
and parallels drawn on the sphere are for guidance only.

\begin{figure}
\centering
\includegraphics[width=0.49\textwidth]{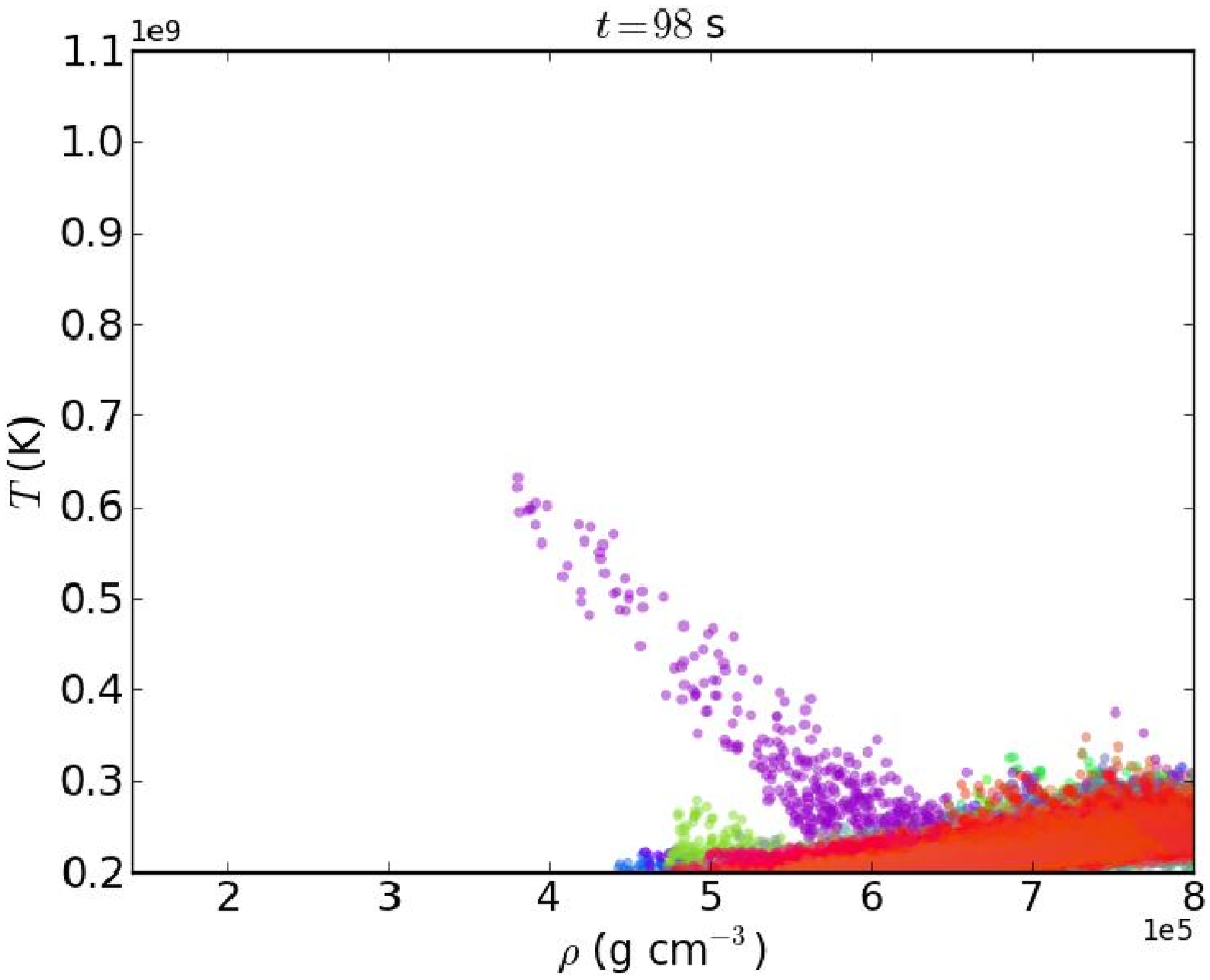}
\includegraphics[width=0.49\textwidth]{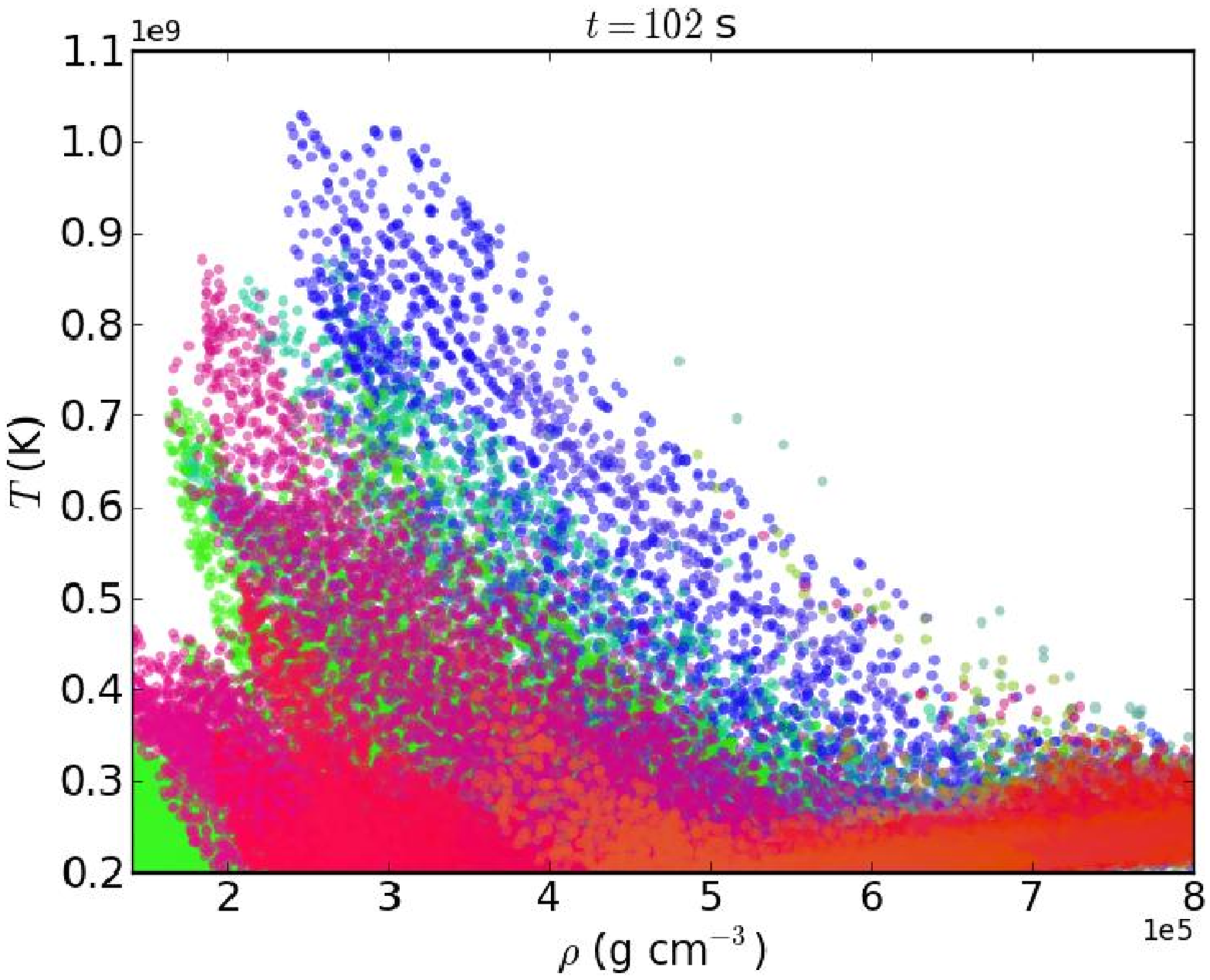}
\caption{\label{fig:rho_t} Distribution in the $\rho-T$ plan of the
  hottest zones in the finest level; the left plot shows a snapshot at
  $t=98$ s and the right plot shows a snapshot at $t=102$ s.  Colors
  represent a proxy for spatial location.  Within four seconds, the
  single dominant plume in the left plot is washed away and replaced
  by several distinct hot regions of which the group of blue zones
  reach runaway conditions.}
\end{figure}

At a single snapshot, it is interesting to see how many hot plumes
exist.  Figure~\ref{fig:rho_t} shows the distribution of the hottest
zones within the $\rho-T$ plane at two different times.  The color
coding of the dots indicate spatial location within the finest level
of refinement on our grid---in particular the normalized $x,y,z$
coordinates of each zone are translated into $r,g,b$ colors for each
data point.  Groups of points of a similar color indicate simulation
zones in close spatial proximity.  The left plot, at $t=98$~s, shows
in purple a single plume dominating the flow.  There are a few smaller
regions beginning to heat, the largest of which is the clump of green
points.  Only four seconds later ($t=102$~s), as seen in the right
plot, the previously dominant plume (purple of the left plot) has
cooled and redistributed its heat while several other distinct plumes
have formed.  This indicates that there are possibly several ignition
regions, or at least several regions that are almost to runaway
conditions once the dominant hot spot (blue points in this case)
ignites.  The extreme temperature sensitivity of the reaction rates
makes it difficult to determine if these other ``failed'' ignition
points would actually ignite.

\begin{figure}
\centering
\includegraphics[width=5in]{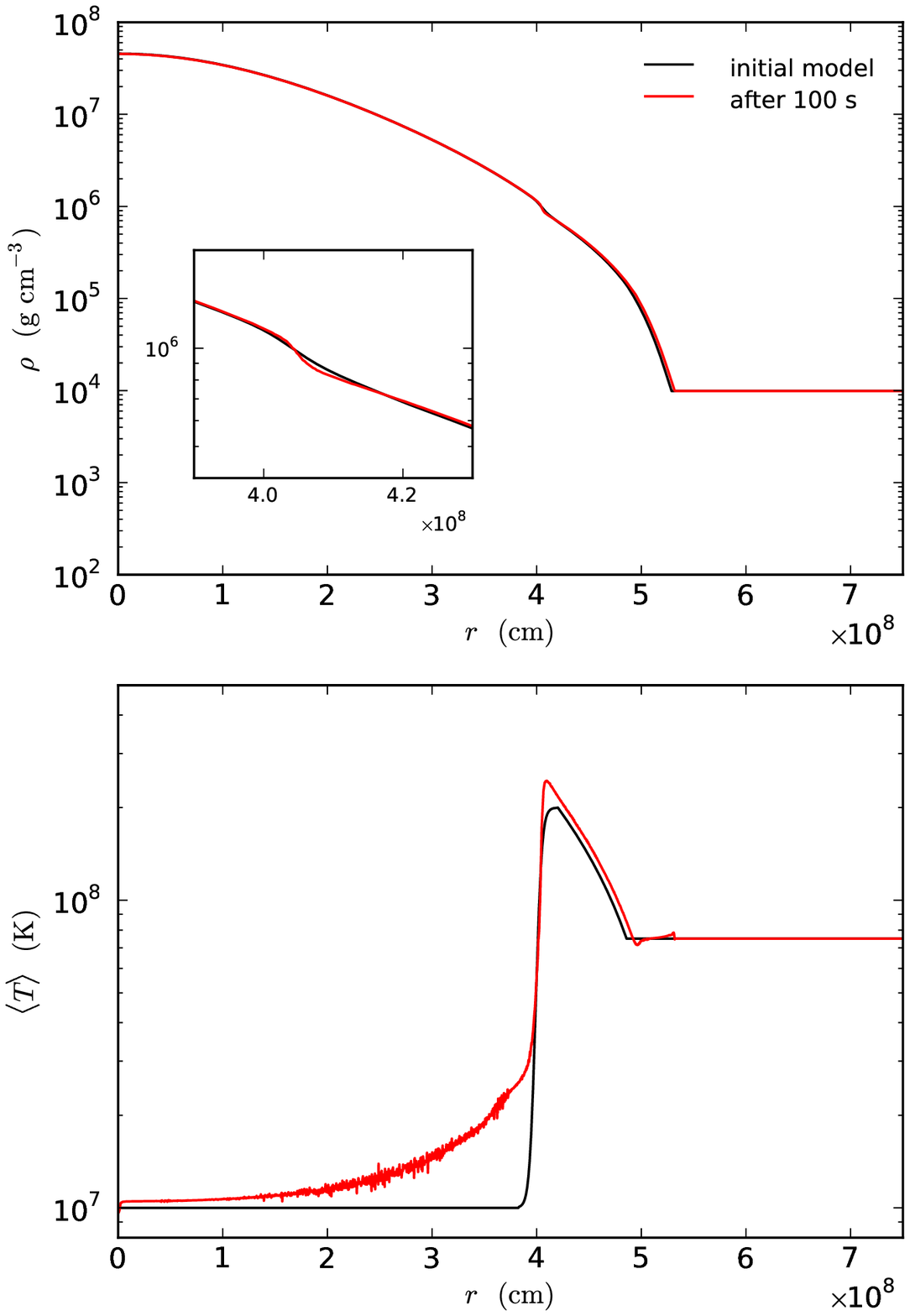}
\caption{\label{fig:base_expand_sim} Base state density and lateral
  average of the temperature at the start and end of the reference
  hot-$\Tcutoff$ simulation.}
\end{figure}

Finally, we can look at how much the atmosphere expanded over the
course of the simulation.  Figure~\ref{fig:base_expand_sim} shows the
base state density at the start of the reference hot-$\Tcutoff$
simulation and after 100~s of evolution, as well as the temperature
averaged in a shell of constant radius at the start and end of the
simulation.  We see that a bit of expansion has taken place, smaller
in magnitude but qualitatively the same as in our one-dimensional
test.  This supports the statement that we need to use the base state
expansion in the hydrodynamics to accurately model the flow.  The
temperature plot shows that the heat generated by reactions has been
transported throughout the convective layer.  We note that there is
heating below the convective layer, which may be due to the adjustment
from the expansion of the base state on the adaptive grid.  The
magnitude of this heating is small and we do not expect it to affect
the results.  This is not seen in the case where we do not burn (next
section).

\subsection{Disabled Burning}\label{sec:disabled burning}

\begin{figure}
\centering
\includegraphics[width=5.0in]{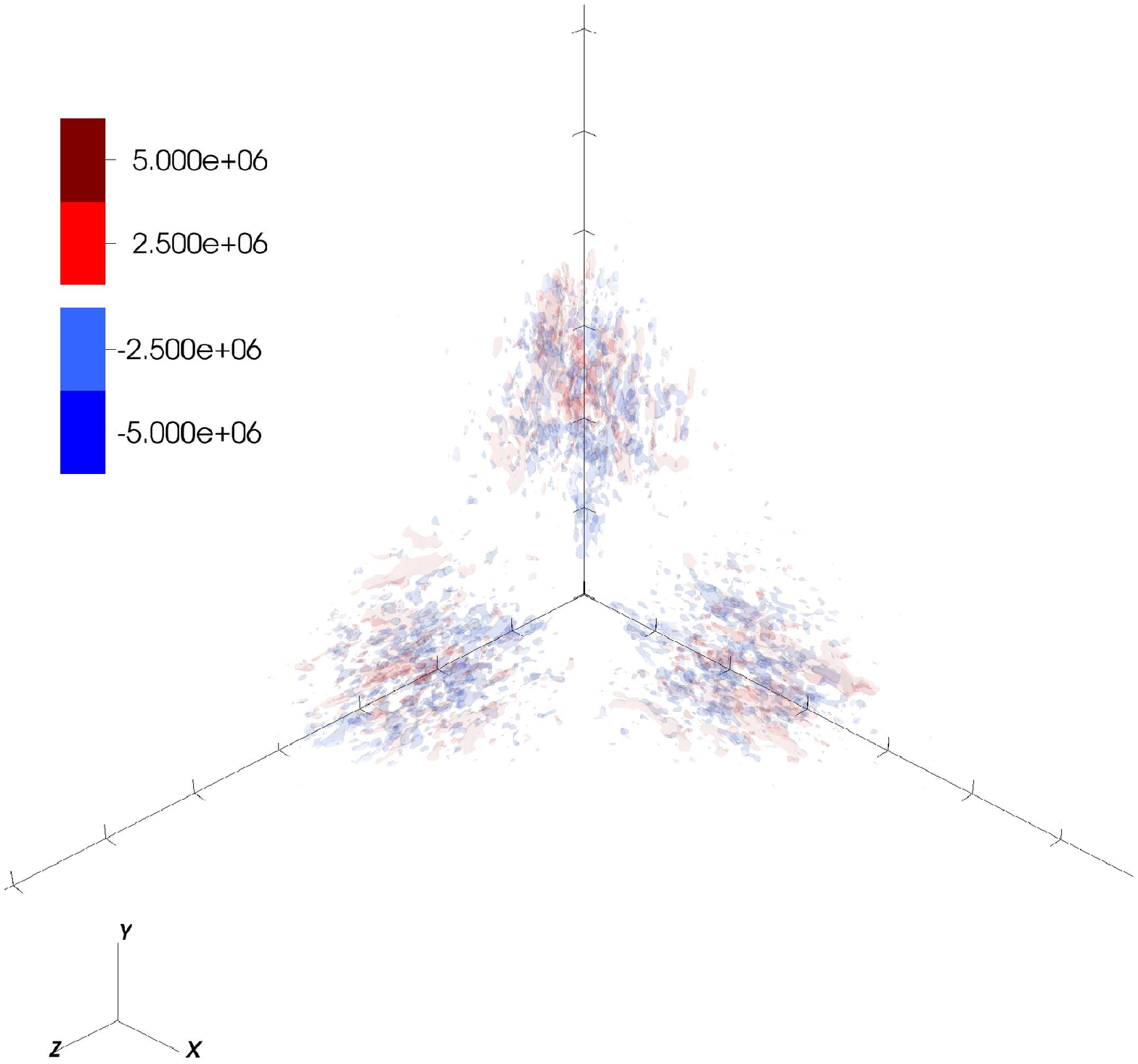} 
\caption{\label{fig:hot_noburn_radvel} Radial velocity (CGS units) for
  the ``no burning'' test model after 197~s.  Note: the range on the
  contours is smaller here than in Figure~\ref{fig:hot_baserun_seq} to
  bring out detail.  The tick marks are $10^8$~cm apart.  We see no
  evidence of the convective field structure.  Instead, the random
  velocities here are driven by discretization error.}
\end{figure}

To demonstrate that the convective velocity field we see is driven by
the reactions, and not due to the initial model's temperature
profile~\citep{mocak:2010} or discretization error, we run a test with
the initial velocity perturbations, but burning disabled.  The radial
velocity field in this case is shown in
Figure~\ref{fig:hot_noburn_radvel}.  We note that the range used in the
contours is smaller than in Figure~\ref{fig:hot_baserun_seq} to bring
out the detail.  We see no suggestion of the convective pattern that
dominates in the burning calculations.  Figure~\ref{fig:baserun}
shows the temperature and Mach number over the course of the
simulations for the reference calculation with burning disabled.  We
see that the peak temperature stays right around the starting value of
$2\times 10^8~\mathrm{K}$, as expected.  The peak Mach number hits a
plateau of $0.05$, driven by the artificial buoyancy introduced by
the mapping error between the
one-dimensional hydrostatic base state and the three-dimensional
Cartesian state.  This value does not seem to grow further.  It is
important to note that a compressible code would also see a velocity
field generated in this test, again driven by the inability to exactly
satisfy hydrostatic equilibrium numerically.  Taken together, these
figures indicate that the convective behavior described above arises
due to the energy release from the reactions.

\subsection{Full Star Simulation}\label{sec:full star}

\begin{figure}
\centering
\plottwo{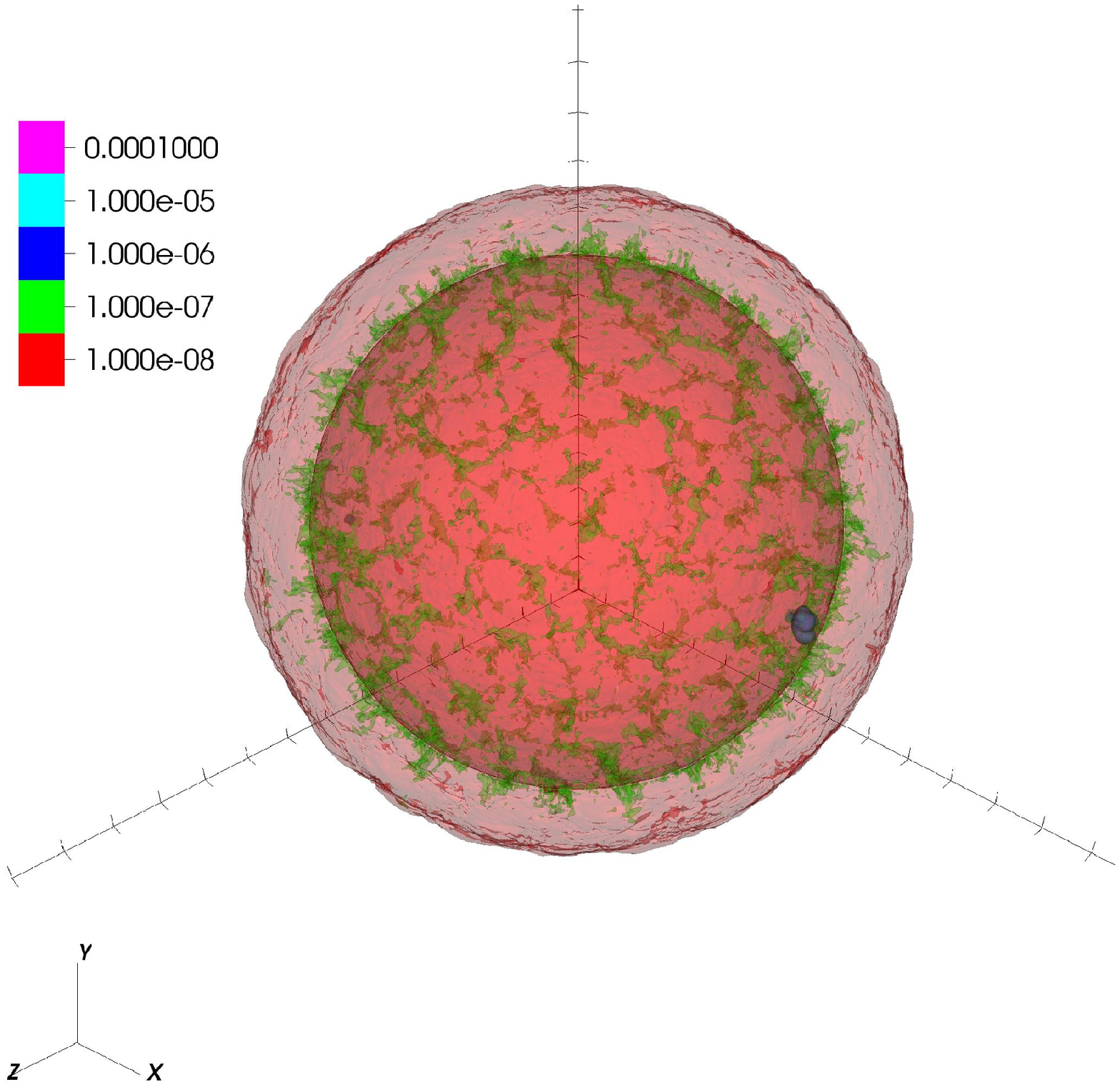}{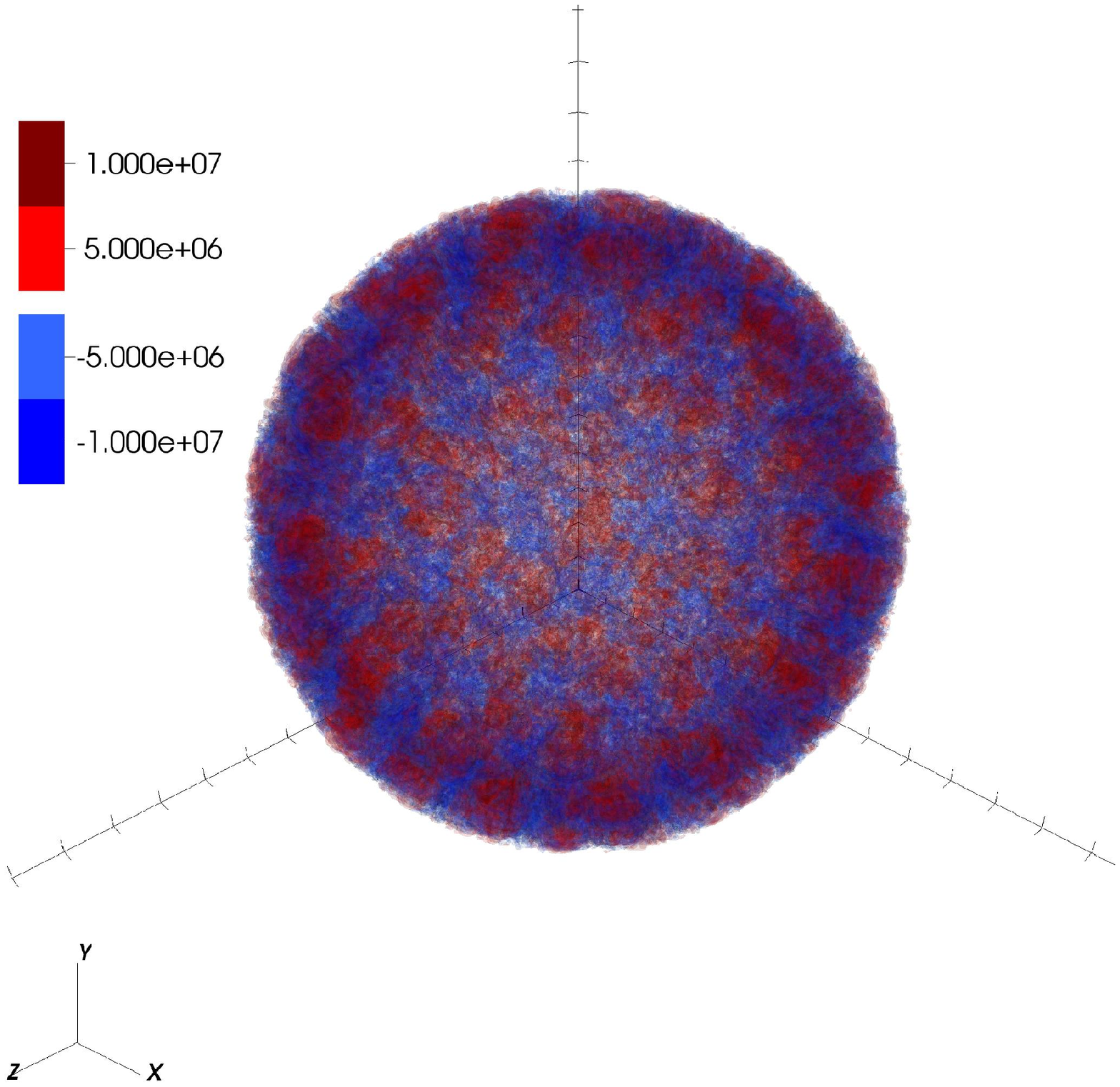}
\caption{\label{fig:hot_fullstar_seq} $^{16}\mathrm{O}$ abundance
  (left) and radial velocity (right; CGS units) for the fullstar
  hot-$\Tcutoff$ model at 96~s---just at the point where we are
  igniting.  The tick marks are $10^8$~cm apart.  The overall
  convective structure compares well to the octant simulations.}
\end{figure}

\begin{figure}
\centering
\includegraphics[scale=0.6]{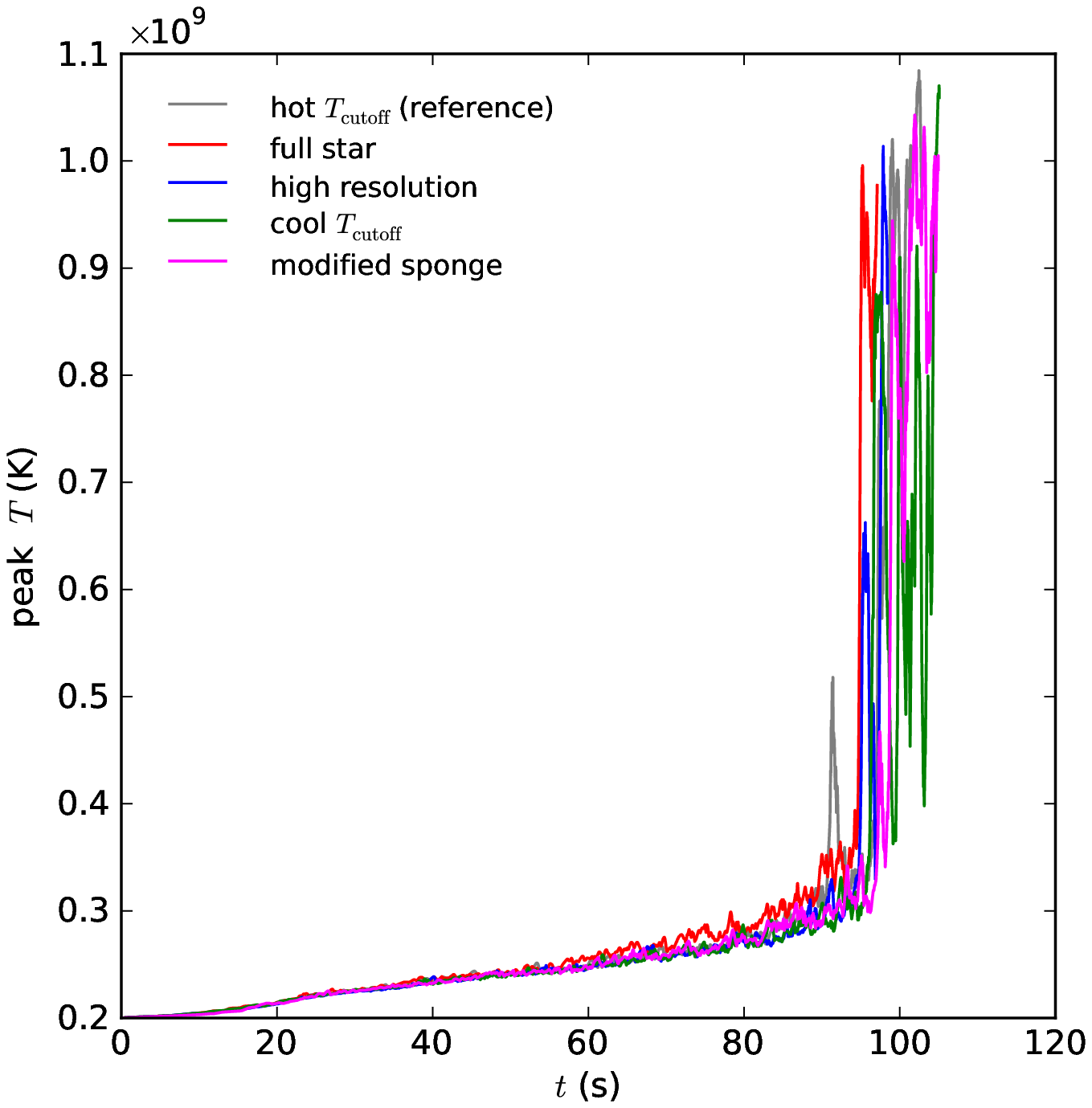} \\
\includegraphics[scale=0.6]{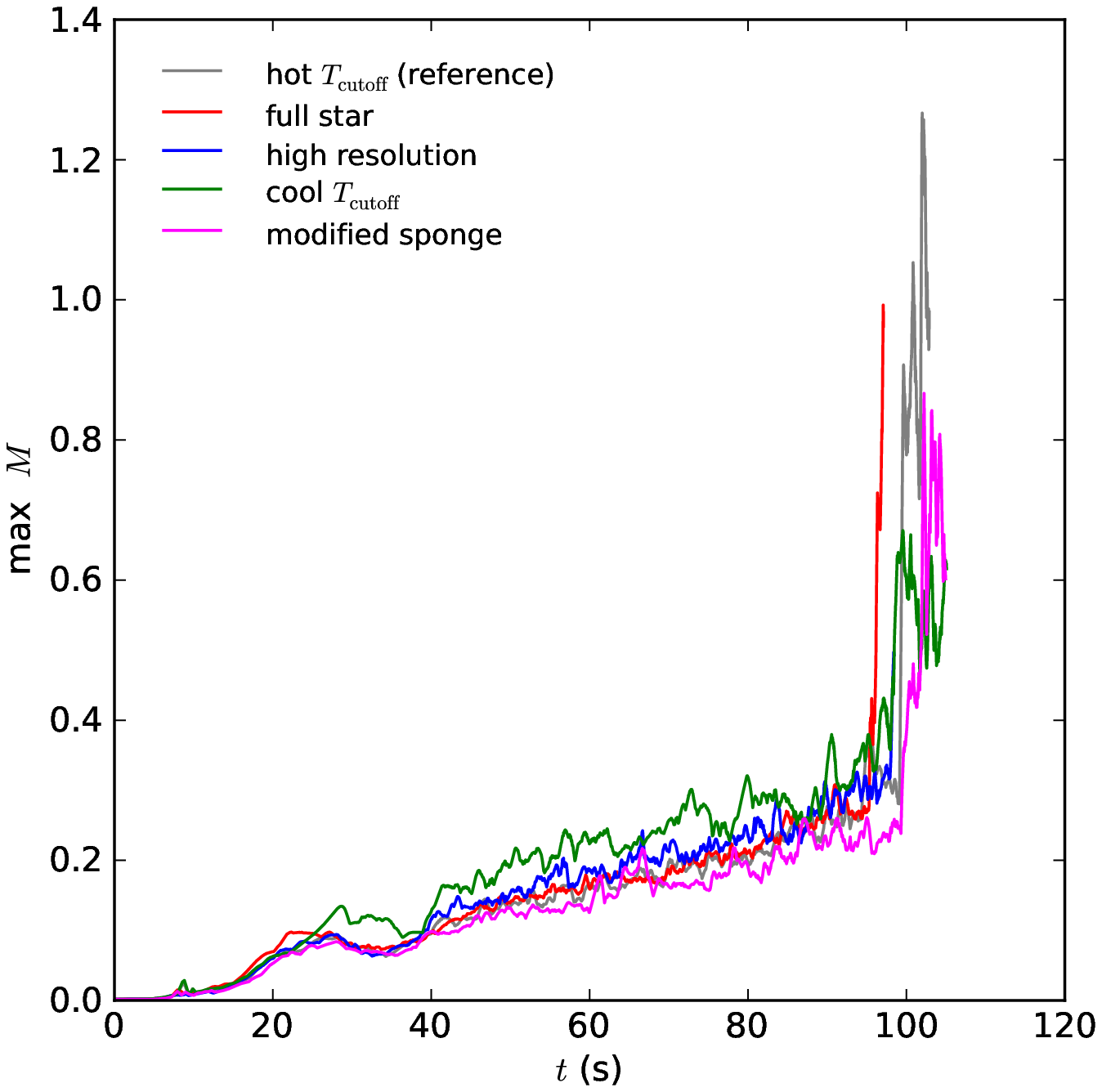} 
\caption{\label{fig:all_trends} Peak temperature and Mach number as a
  function of time for the various supporting calculations: full star,
  high-resolution, cool-$\Tcutoff$ model, and stronger sponging.  The
  reference hot-$\Tcutoff$ calculation is also shown.  We see that the
  trends for all quantities are the same for all runs, indicating that
  our simulation is converged in resolution, and the octant is a
  reasonable model for the energetic evolution, and that the treatment
  at the outer radius of the convective region is not critically
  important.}
\end{figure}

\begin{figure}
\centering
\includegraphics[width=5in]{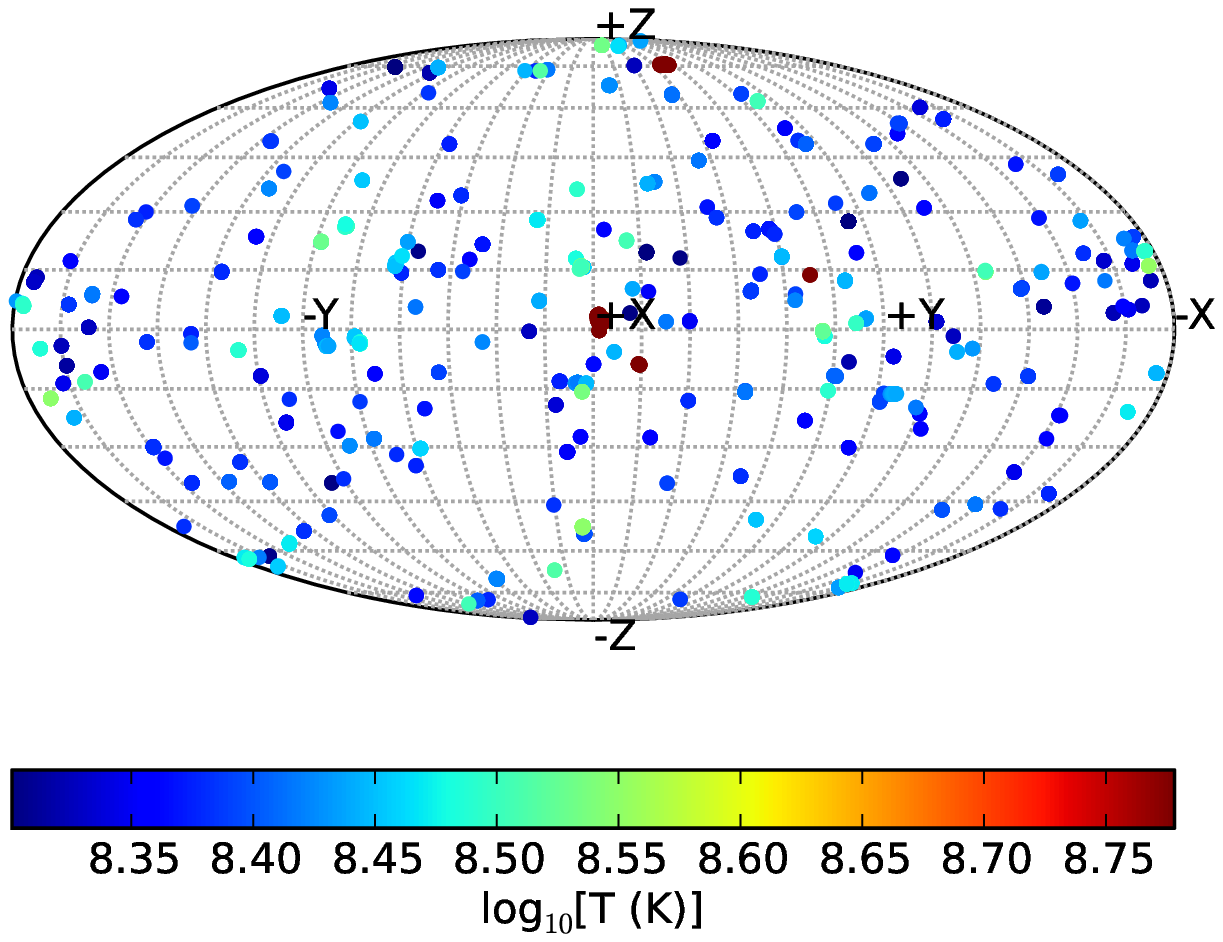}
\caption{\label{fig:fullstar_hotspot} Location of the hottest point at
  each time step for the full-star, hot-$\Tcutoff$ calculation.}
\end{figure}

We run a single full-star calculation (using the hot-$\Tcutoff$ model)
to assess the influence of the octant geometry on the general results
described above.  The resolution is the same, with the base grid now
twice as large in each coordinate direction.  The convective field is
shown in Figure~\ref{fig:hot_fullstar_seq}, and we see the same
overall structure that appears in the octant simulations.  The
time-dependent peak temperature and Mach number
(Figure~\ref{fig:all_trends}) also agree well with the
octant case.  Figure~\ref{fig:fullstar_hotspot} shows the hotspot
location over time for this calculation.  We see a uniform
distribution of points over the sphere. 

\subsection{High-Resolution}\label{sec:resolution sensitivity}

\begin{figure}
\centering
\plottwo{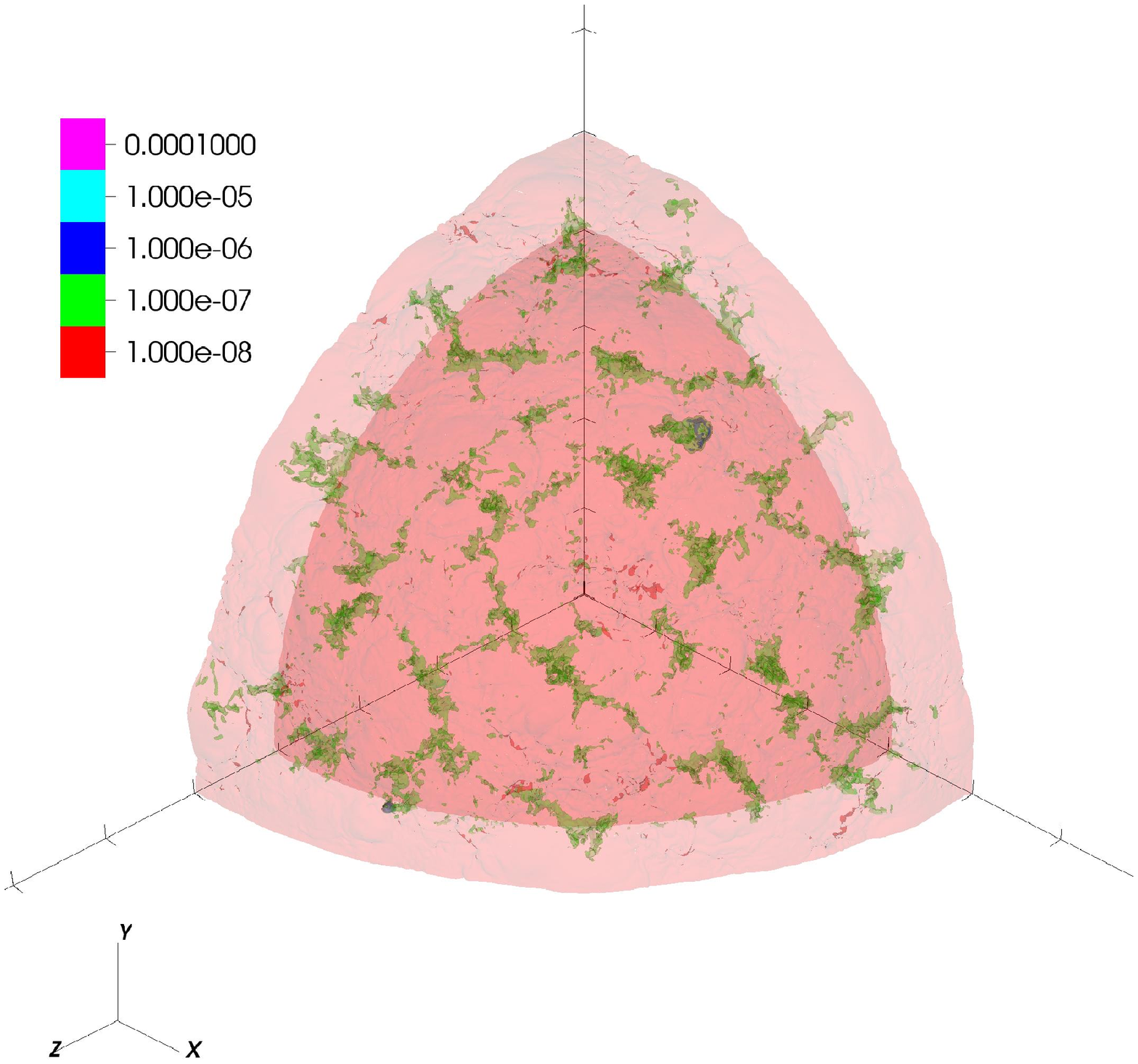}{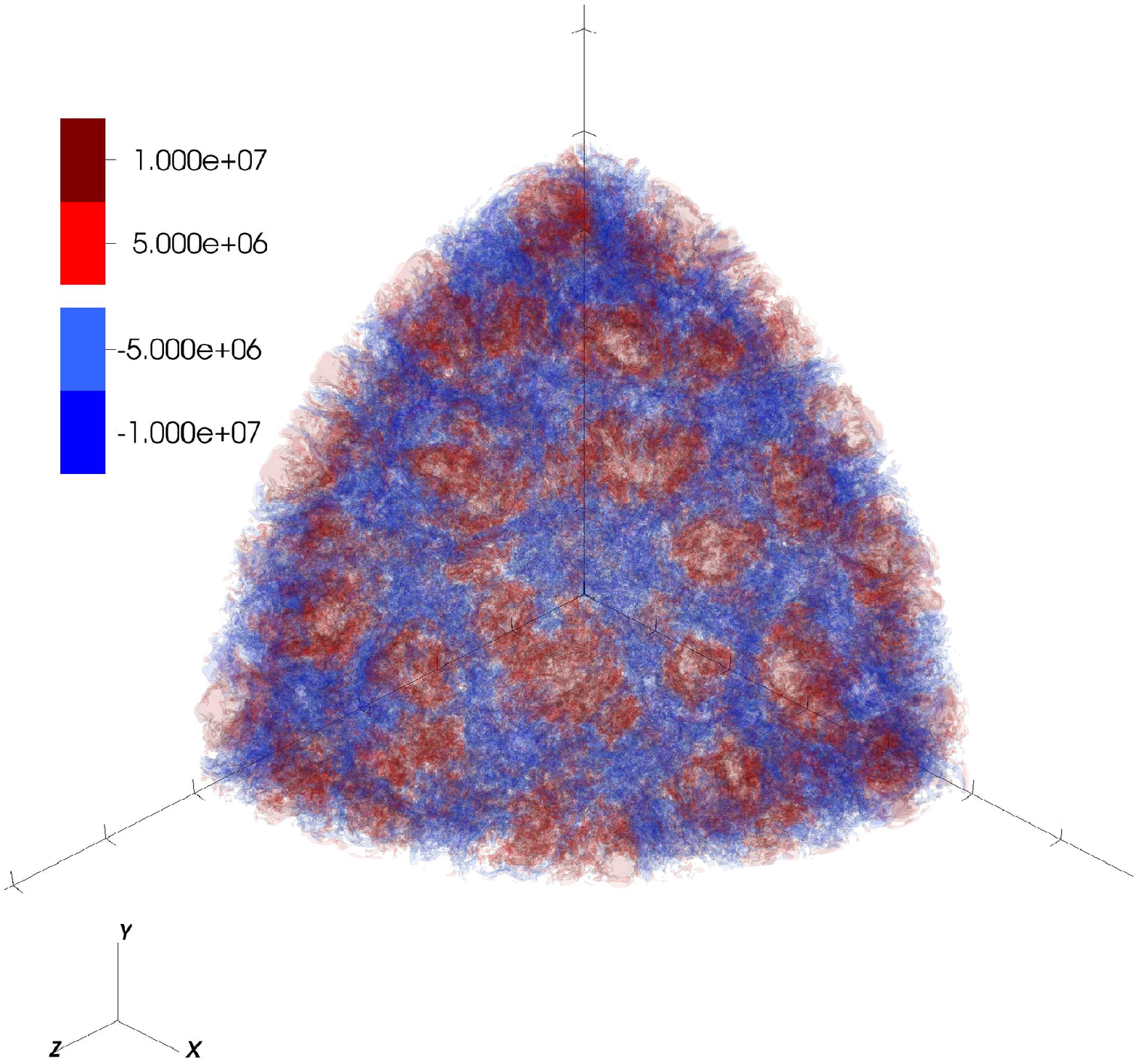}
\caption{\label{fig:hires_seq} $^{16}\mathrm{O}$ abundance (left) and
  radial velocity (right; CGS units) for the high-resolution
  hot-$\Tcutoff$ near the point of ignition (98~s).  The tick marks are
  $10^8$~cm apart.  The overall
  structure compares well to the standard-resolution hot-$\Tcutoff$
  simulation.}
\end{figure}

To understand the robustness of the convective features to
resolution, we run a single case (hot-$\Tcutoff$ model) at twice the 
resolution.  This is accomplished by doubling the number of zones 
in each direction on the coarse grid.
Figure~\ref{fig:hires_seq} shows the convective field for this
simulation, in which the overall structure agrees with that of the
standard-resolution simulations. The time-dependent peak temperature and 
Mach number for this run (Figure~\ref{fig:all_trends}) again show excellent 
agreement with the standard-resolution runs.  This gives us confidence
that we are sufficiently numerically converged.

\subsection{Cooler Initial Model}\label{sec:cool model}

\begin{figure}
\centering
\plottwo{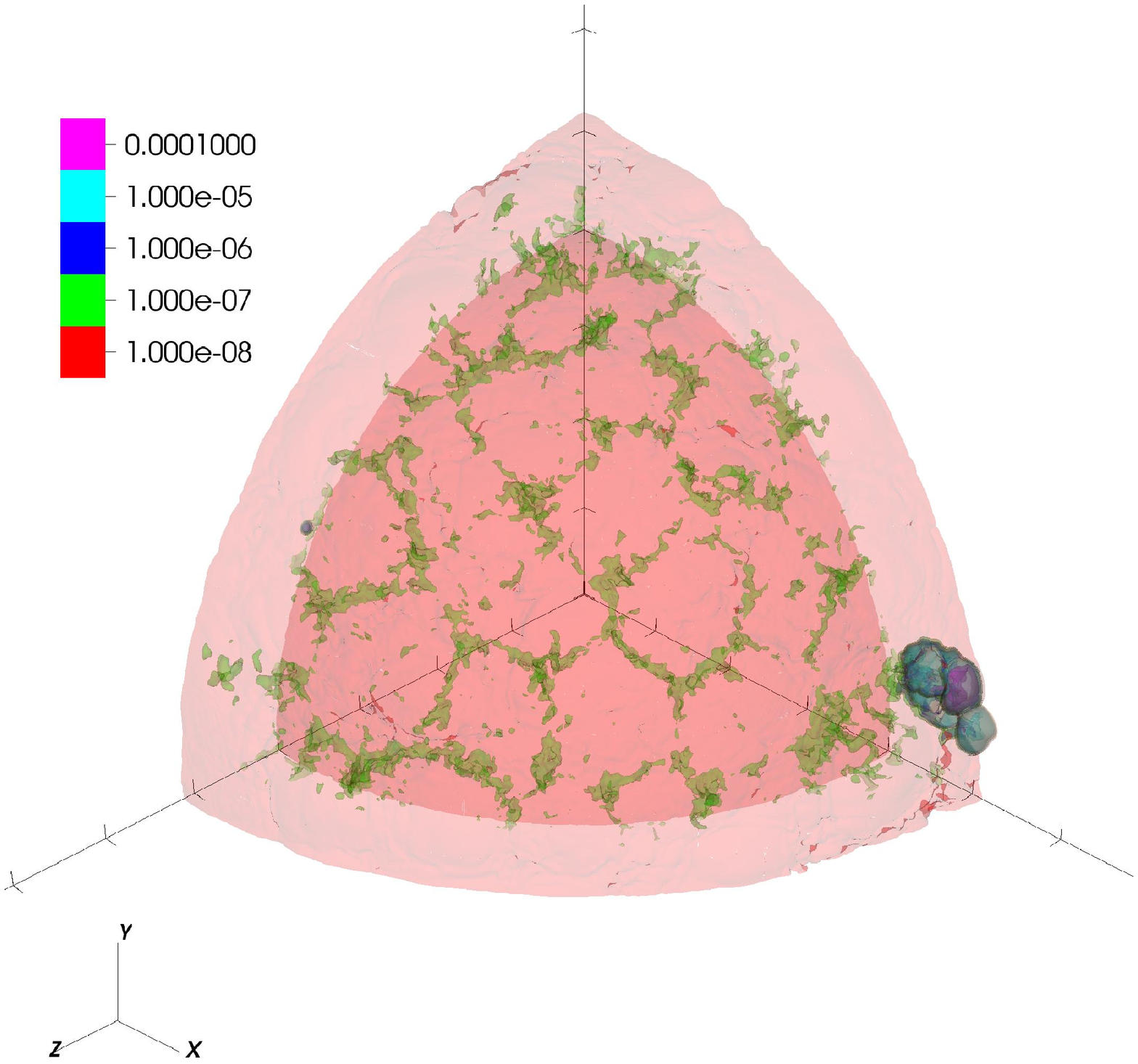}{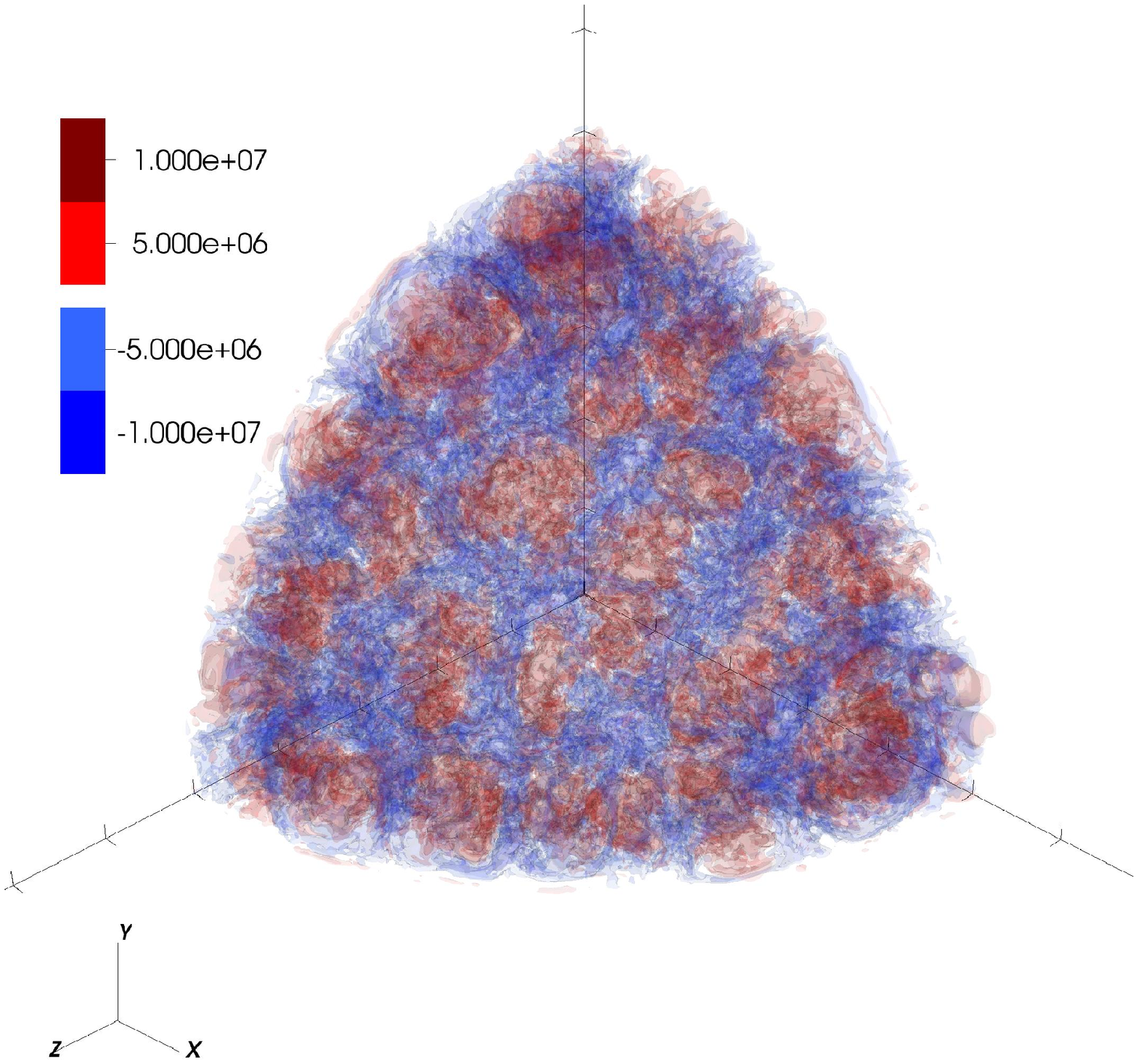}
\caption{\label{fig:cool_baserun_seq} $^{16}\mathrm{O}$ abundance
  (left) and radial velocity (right; CGS units) for the
  cool-$\Tcutoff$ model at 100~s.  The tick marks are $10^8$~cm apart.
  The overall structure compares well to the hot-$\Tcutoff$
  simulation.}
\end{figure}

The corresponding images for the cool-$\Tcutoff$ simulation near the
point of ignition are shown in Figure~\ref{fig:cool_baserun_seq}.
There is a strong qualitative similarity with the hot-$\Tcutoff$ run,
indicating that the structure of the convection is insensitive to the
details of the top of the convective zone.
We also see the expected behavior that the hot-$\Tcutoff$ model has a
lower peak $M$ than the cool-$\Tcutoff$ model because of the higher
sound speed at the edge of the star.  This suggests that the peak Mach
number occurs at the edge of the star, outside of the convective
region.

\subsection{Effect of Cutoff Densities and Sponging}\label{sec:sponge}

In the simulations above, we chose the sponging parameters such that
the sponging begins just at the top of the convective layer.  Here we
explore the sensitivity of that choice by turning the sponge on
lower---we now set $\rho_\mathrm{md} = \rho_\mathrm{anelastic} =
10^5~\gcc$.  Additionally, we decrease the timescale over which the
sponging acts by changing $\kappa = 10~\mathrm{s^{-1}}$ to 
$\kappa = 100~\mathrm{s^{-1}}$ (see \citealt{ABNZ:III}).

\begin{figure}
\centering
\plottwo{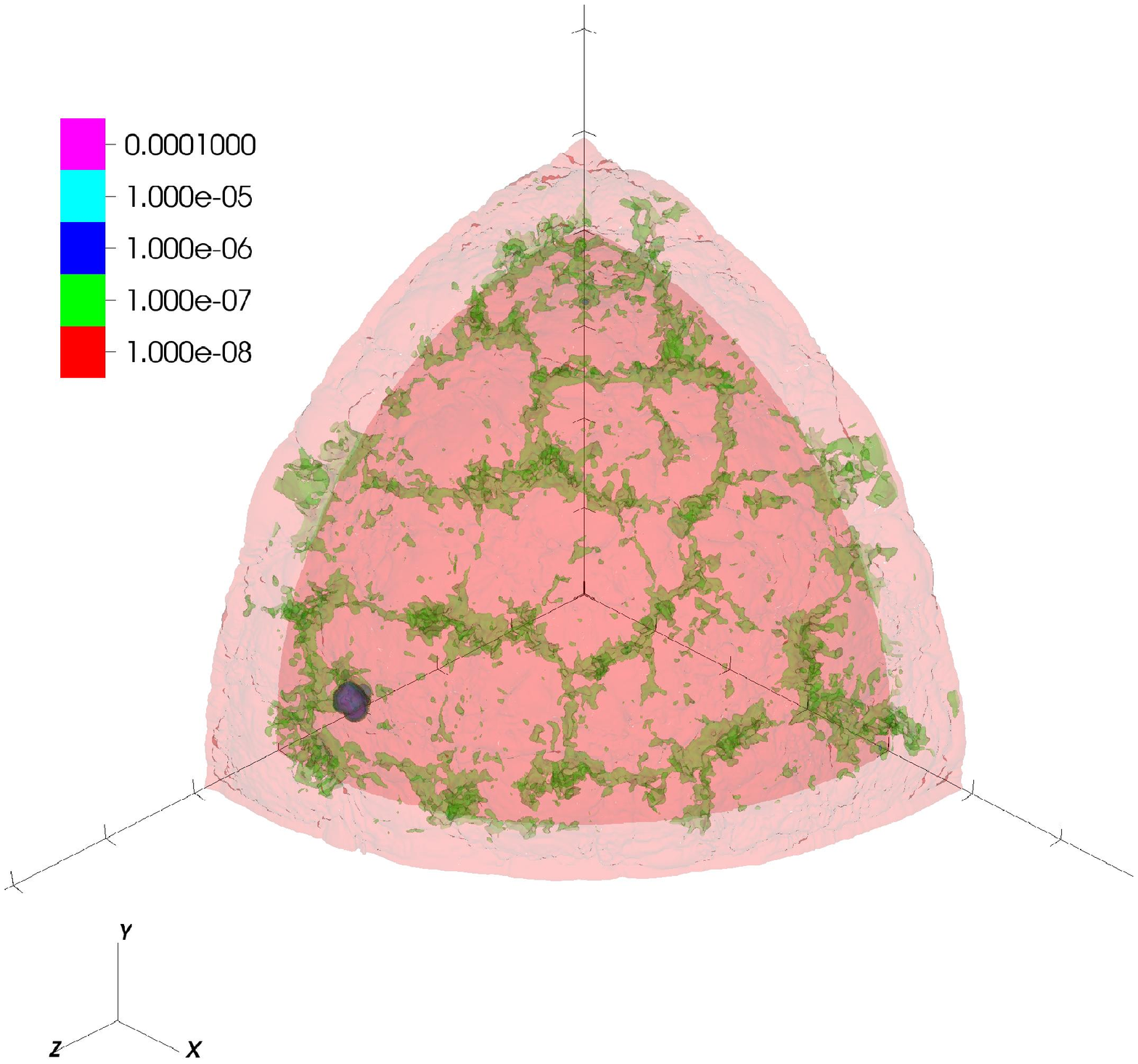}{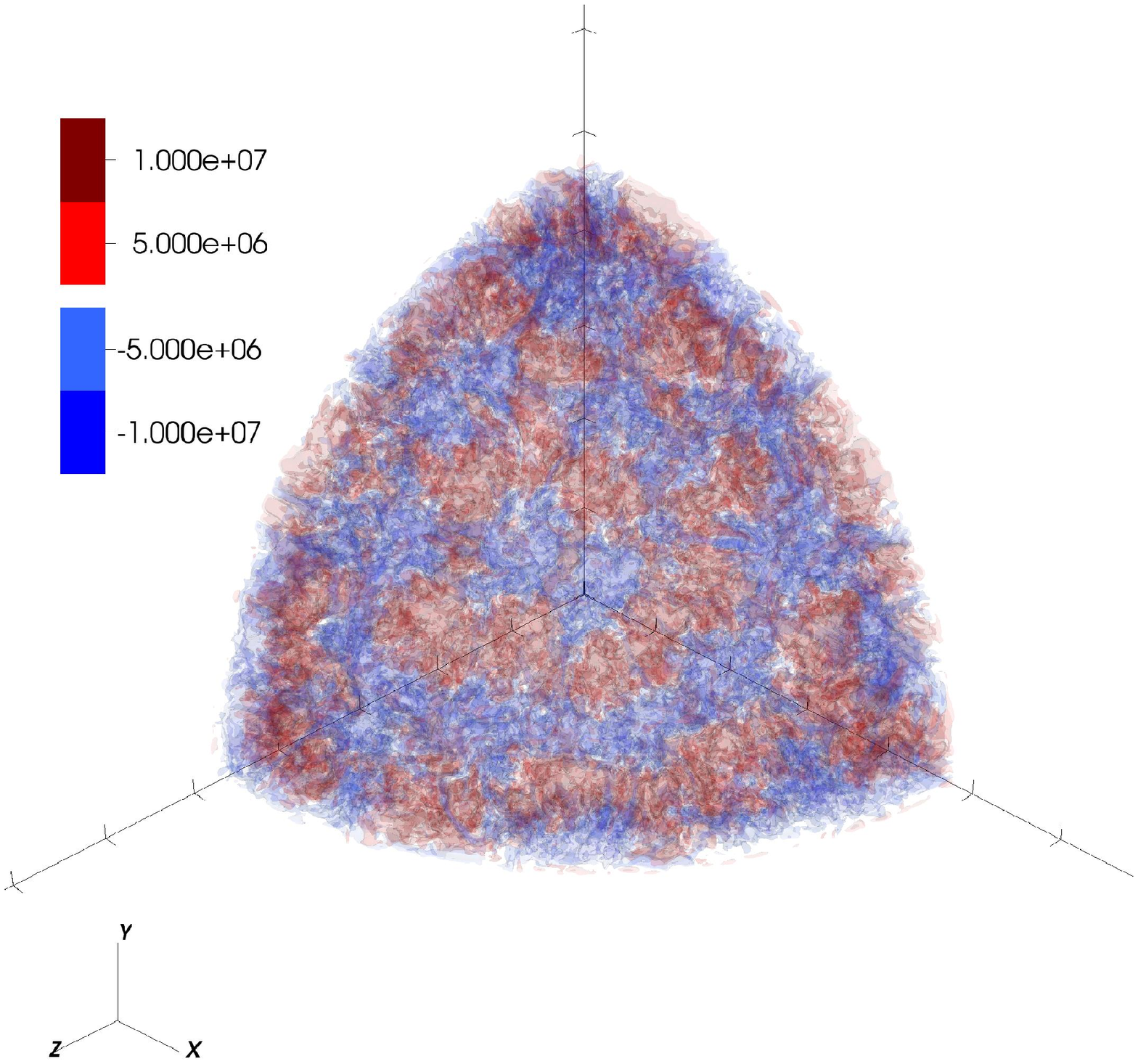}
\caption{\label{fig:sponge_seq} $^{16}\mathrm{O}$ abundance (left) and
  radial velocity (right) for the hot-$\Tcutoff$ run with more
  aggressive sponging at the top of the convective layer.  Shown at
  100~s.  The tick marks are $10^8$~cm apart.  Again, we see good
  agreement with the other cases.}
\end{figure}

Figure~\ref{fig:sponge_seq} shows the structure of the convective
field.  We notice that the radial extent appears slightly diminished
compared to the previous simulations, owing to the more aggressive
sponging.  However, the overall pattern of convective cells appears
consistent with the other cases.   The trends of peak temperature and Mach number are shown
in Figure~\ref{fig:all_trends}, and exhibit a slightly lower peak Mach
number as the reference simulation due to the more aggressive sponging.
These comparisons show that the convective
behavior is not strongly dependent on how we treat the top of the
convective layer.

\section{Conclusions and Discussion}\label{Sec:Conclusions and Discussion}

The main goal of the present paper is to serve as a proof-of-concept that the low
Mach number methodology can be applied to shell burning on the surface
of white dwarfs.  This is the first application of MAESTRO where we have
had off-center heating with an expanding, self-gravitating hydrostatic
state.  We demonstrated that efficient three-dimensional models of the
convective flow leading up to the ignition of a burning front in a
helium layer on the surface of a white dwarf are possible.  We
explored the sensitivity of our results to a variety of factors and
showed that the convective features realized are robust.  The octant
calculations are inexpensive to run (requiring only about 40,000 CPU
hours on the OLCF jaguar machine, using 128 MPI tasks with 8 OpenMP
threads per task).  This suggests that a parameter study of progenitor
models is feasible---this will be the focus of a follow-on paper.
There is a wide variety of potential models---varying white dwarf
masses and helium envelopes.  The conditions at the base of the helium
layer will vary across these different models, so some models may be
more amenable to our methodology than others.  This needs to be
explored on a case-by-case basis.  For instance, we expect that low
mass shells and lower mass white dwarfs would have slower dynamics.

Future work will focus on better understanding the conditions
in the helium layer leading up to ignition.  Open scientific questions
that we wish to understand are whether the ignition occurs in a manner
that is amenable to a detonation---this is a key requirement for the
sub-Chandrasekhar explosion models.  Also of interest is whether
ignition can arise in multiple disjoint points on the surface of the
white dwarf.  Studying this will require either enhancements to the
low Mach number model (i.e.\ the addition of long wavelength
acoustics) or feeding MAESTRO models into a compressible hydrodynamics
code.  Finally, the models presented here started rather late in the
evolution, but it is easy to start with cooler models to see more of
the ramp up to ignition.  This earlier evolution will take place at
lower Mach numbers.

\acknowledgments

Videos of the reference calculation are available at:
\url{http://youtu.be/boHVbcfazvw} and \url{http://youtu.be/37WqQOKm0p4}.
We thank Frank Timmes for making his equation of state routines
publicly available and for helpful discussions on the thermodynamics.
The work at Stony Brook was supported by a DOE/Office of Nuclear
Physics grant No.~DE-FG02-06ER41448 to Stony Brook.  The work at LBNL
was supported by the Applied Mathematics Program of the DOE Office of
Advance Scientific Computing Research under U.S. Department of Energy
under contract No.~DE-AC02-05CH11231.

An award of computer time was provided by the Innovative and Novel
Computational Impact on Theory and Experiment (INCITE) program.  This
research used resources of the Oak Ridge Leadership Computing Facility
located in the Oak Ridge National Laboratory, which is supported by
the Office of Science of the Department of Energy under Contract
DE-AC05-00OR22725.  Visualizations were performed using VisIt
and matplotlib.

\clearpage


\clearpage

\clearpage

\clearpage

\begin{deluxetable}{lll}
\tablecolumns{3}
\tablewidth{0pt}
\tablecaption{\label{table:cutoffs} Cutoff parameter values}
\tablehead{\colhead{parameter} & 
           \colhead{cool-$\Tcutoff$ model} &
           \colhead{hot-$\Tcutoff$ model}}
\startdata
$\rhocutoff$        &  $10^4~\gcc$         &  $10^4~\gcc$ \\
$\rho_\mathrm{md}$  &  $3\times 10^4~\gcc$ &  $6\times 10^4~\gcc$ \\
$f_\mathrm{sp}$     &  $2.0$               &  $2.0$ \\
$\rhoanelastic$     &  $3\times 10^4~\gcc$ &  $6\times 10^4~\gcc$ \\
\enddata
\end{deluxetable}  

\clearpage

\clearpage

\clearpage

\begin{figure}
\centering
\figurenum{\ref{fig:hot_baserun_seq}}
\includegraphics[scale=0.3]{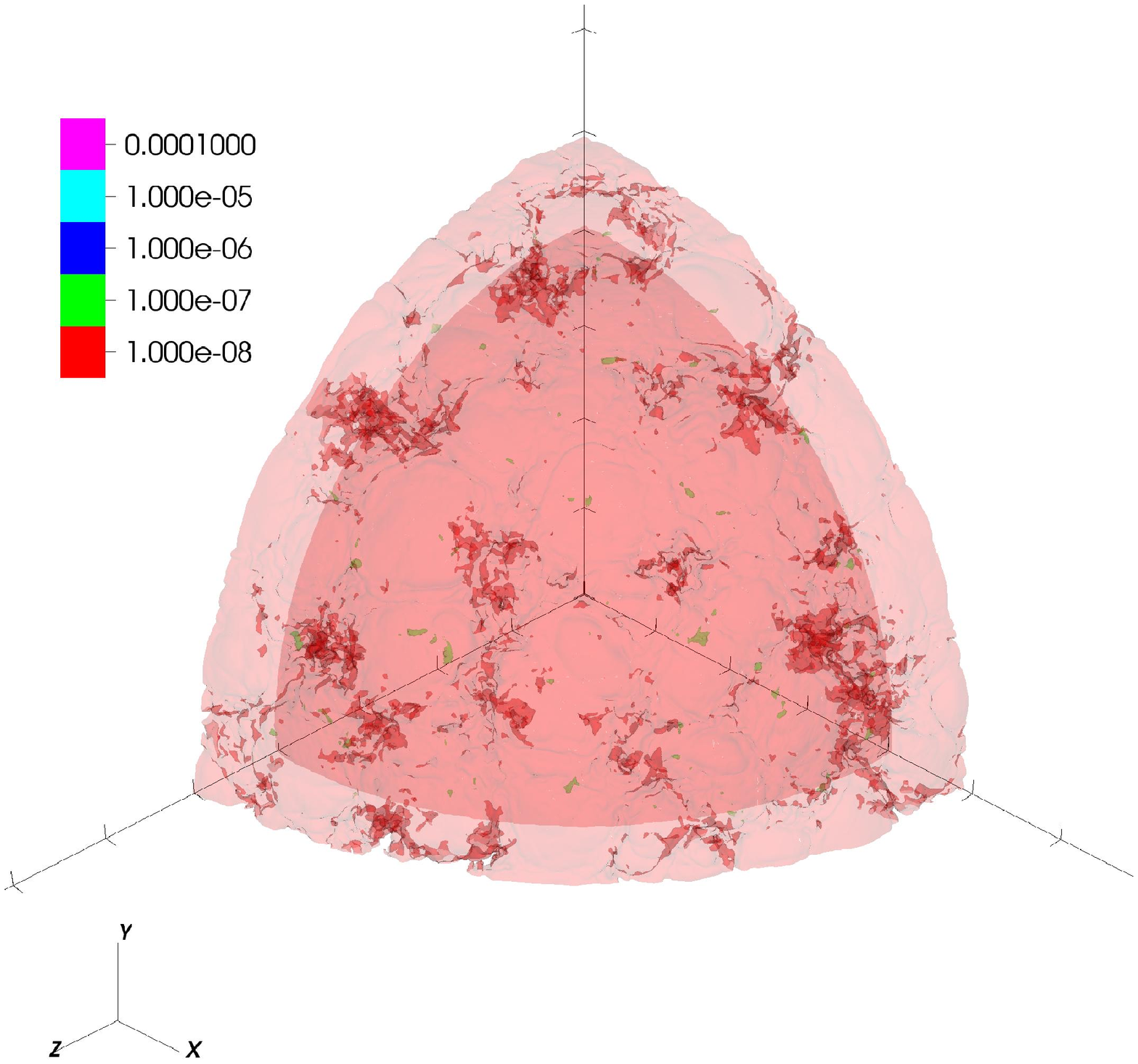}\hspace{1em}
\includegraphics[scale=0.3]{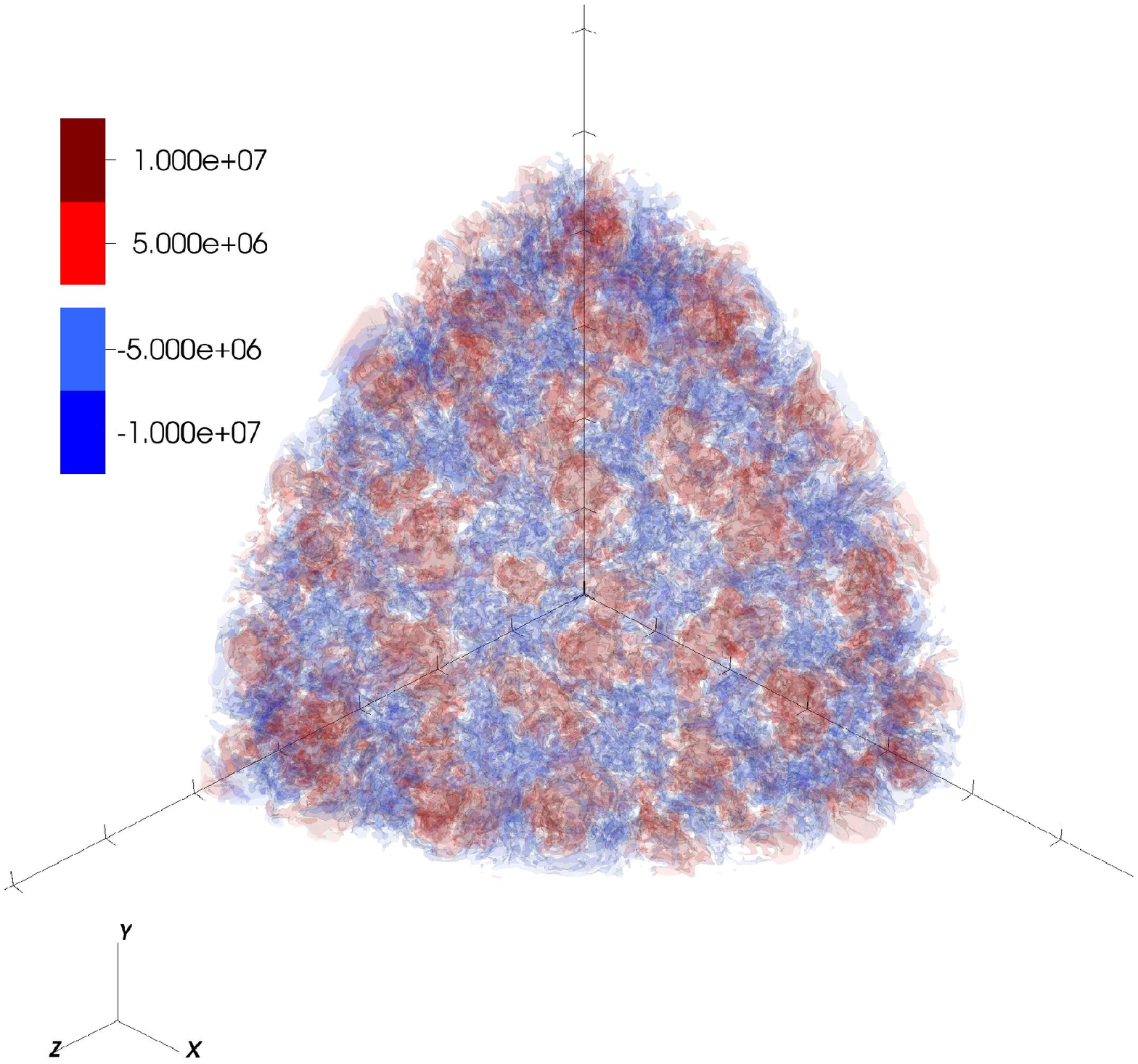} \\
\includegraphics[scale=0.3]{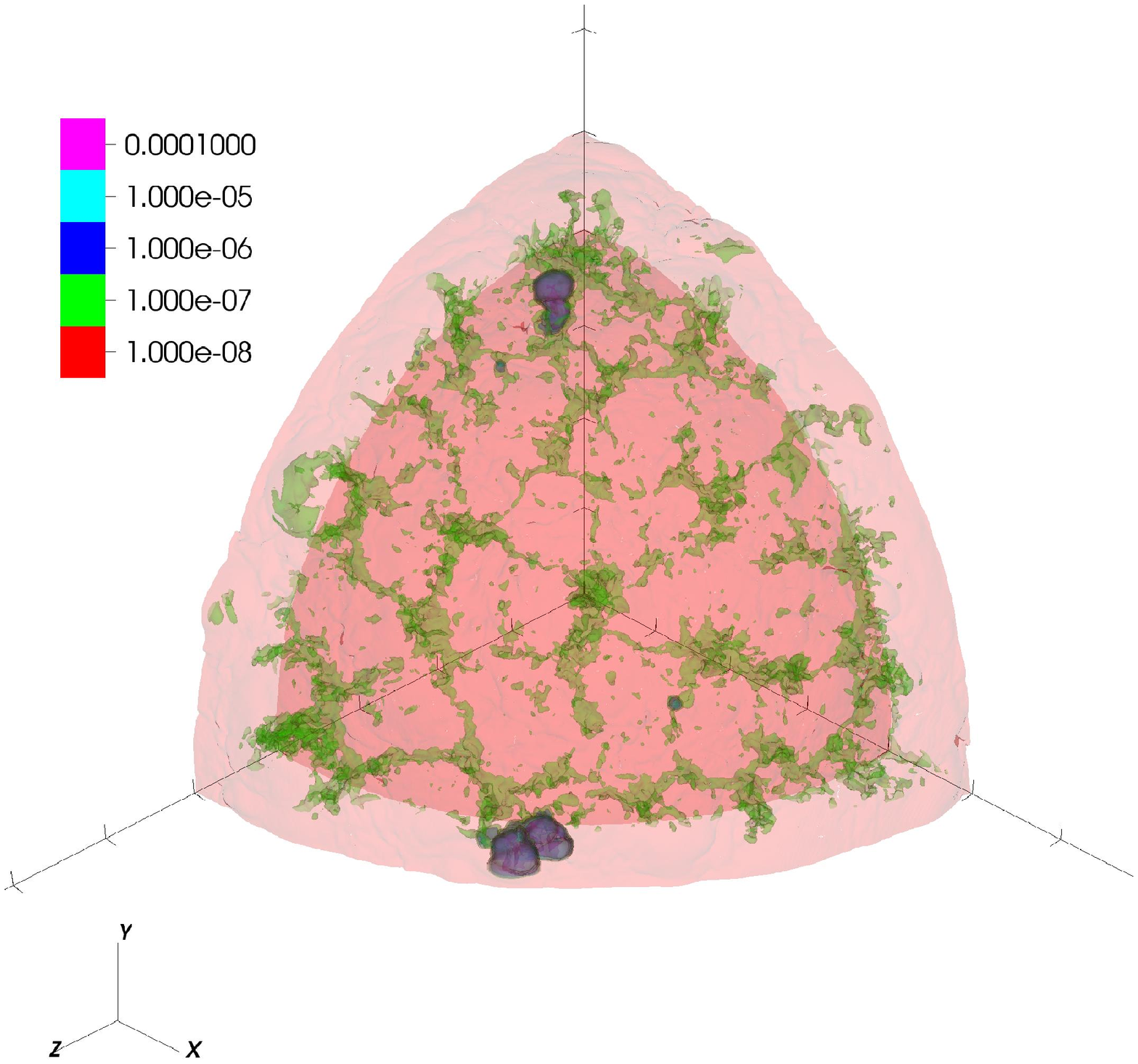}\hspace{1em}
\includegraphics[scale=0.3]{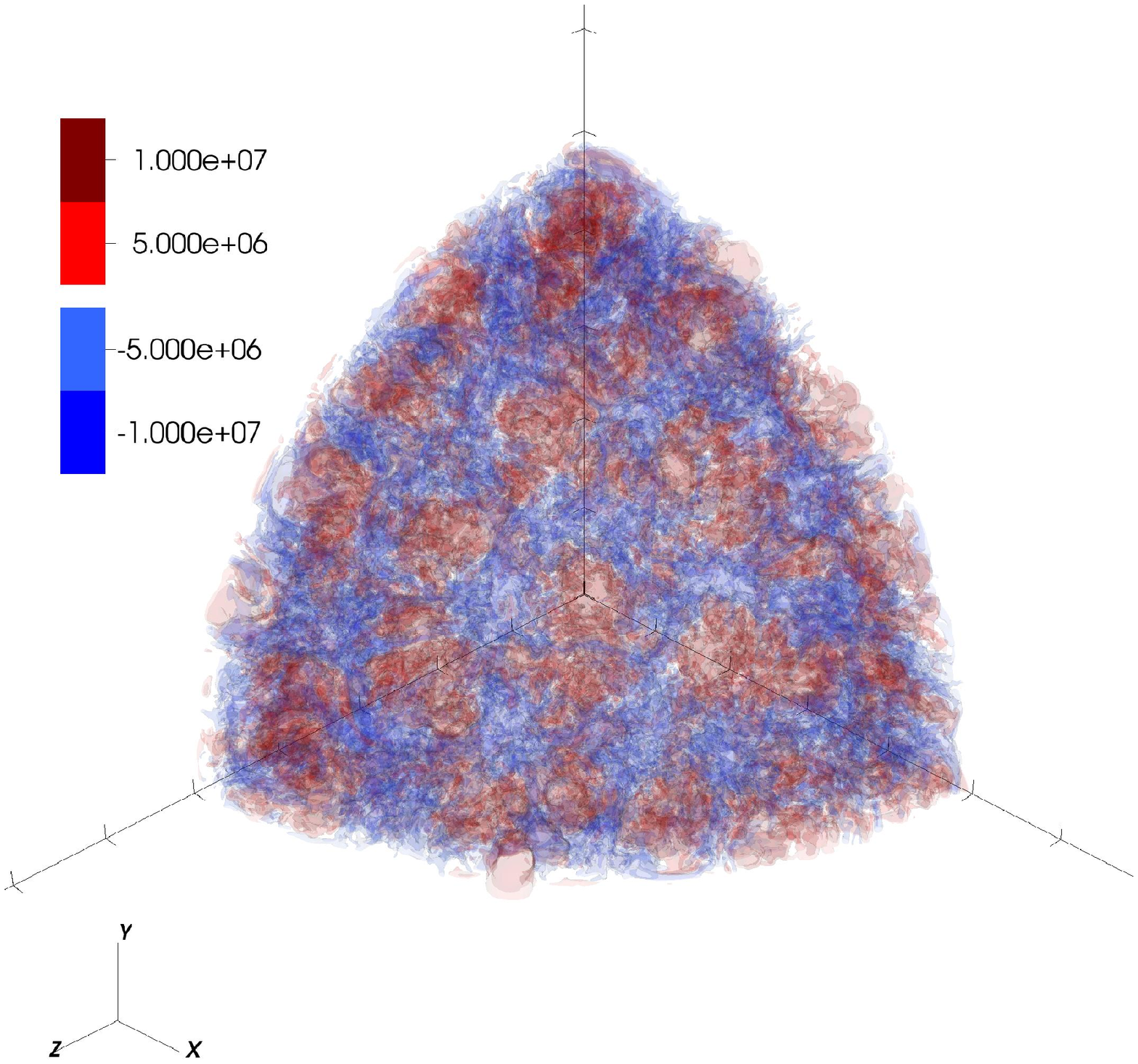} 
\caption{Time sequence continued.}
\end{figure}

\clearpage

\clearpage

\clearpage

\clearpage

\clearpage

\clearpage

\clearpage

\clearpage

\clearpage

\clearpage

\clearpage

\clearpage

\end{document}